\documentstyle[aps,prb,array,psfig,multicol]{revtex}
\begin{document}

\title{Dynamical response of a pinned two-dimensional Wigner
crystal}

\author{Michael M. Fogler}

\address{School of Natural Sciences, Institute for Advanced Study,
Einstein Drive, Princeton, New Jersey 08540}

\author{David A. Huse}

\address{Physics Department, Princeton University,
Princeton, New Jersey 08544}

\maketitle

\raisebox{125pt}[0pt][0pt]{
\mbox{\hspace{4.95in}Preprint IASSNS-HEP-00/32}
}

\vspace{-0.3in}

\begin{abstract}

We re-examine a long-standing problem of a finite-frequency
conductivity of a weakly pinned two-dimensional classical Wigner
crystal. In this system an inhomogeneously broadened absorption line
(pinning mode) centered at disorder and magnetic field dependent
frequency $\omega_p$ is known to appear. We show that the relative
linewidth $\Delta \omega_p / \omega_p$ of the pinning mode is of the
order of one in weak magnetic fields, exhibits a power-law decrease in
intermediate fields, and eventually saturates at a small value in
strong magnetic fields. The linewidth narrowing is due to a peculiar
mechanism of mixing between the stiffer longitudinal and the softer
transverse components of the collective excitations. The width of the
high-field resonance proves to be related to the density of states in
the low-frequency tail of the zero-field phonon spectrum. We find a
qualitative agreement with recent experiments and point out
differences from the previous theoretical work on the subject.

\end{abstract}
\pacs{PACS numbers: 73.40.Hm, 75.40.Gb, 63.50.+x, 73.20.Mf}

\begin{multicols}{2}

\section{Introduction}
\label{Introduction}

The problem of pinning of an elastic solid by an external random
potential has been a long-standing problem of condensed matter
physics, first addressed by Larkin~\cite{Larkin_70} in 1970. Physical
systems where this problem arises include charge-density
waves,~\cite{Gruner_88} vortex lattices in
superconductors,~\cite{Blatter_94} interfaces,~\cite{Toner_90}
magnetic bubbles,~\cite{Seshadri_92} liquid crystals in
aerogels,~\cite{Radzihovsky_99} and many others. An interesting
example of the charge-density wave is a Wigner crystal (WC). Under
terrestrial conditions a stable WC has been realized in its
two-dimensional (2D) form,~\cite{3D_WC} on cryogenic surfaces and in
semiconductor heterostructures with low carrier density and/or in
strong magnetic fields.~\cite{WC_books} In this work we examine the
finite-frequency response of a collectively pinned 2D WC. The
collective pinning is the regime where individual defects of the
substrate are too weak to significantly deform the crystalline
lattice, so that the WC is well ordered (``unpinned'') at small length
scales, but the cumulative effect of the disorder eventually dominates
the elasticity at a pinning length, which is much larger than the
lattice constant.

The linear-response conductivity $\sigma_{\alpha\beta}(q, \omega)$ of
the WC in an external magnetic field is a tensor. We focus our study on
the real part of its diagonal component $\sigma_{xx}(q = 0, \omega)$,
the quantity that determines the power absorption in the presence of a
uniform electric field. If the frequency $\omega$ is not too small, the
absorption is dominated by the collective excitations. In the absence of
the magnetic field and random potential, a 2D WC possesses two branches
of such excitations: the longitudinal (L) and the transverse (T)
phonons. The long-wavelength L-phonon is also referred to as plasmon. In
a finite magnetic field L- and T-phonons hybridize into the
magnetophonons (gapless lower hybrid mode) and the magnetoplasmons
(gapped higher hybrid mode). It is noteworthy that without the pinning
potential, the frequency and the oscillator strength of the
magnetophonons vanish at $q = 0$ by virtue of Kohn's theorem. The
absorption of a spacially uniform ac field is possible only by exciting
the magnetoplasmon mode, at the cyclotron frequency $\omega_c = e B /
m_e c$ set by the magnetic field $B$. The pinning shifts the original $q
= 0$ magnetophonon mode to a nonzero disorder-dependent frequency
$\omega_p$, broadens it, and imparts it with a finite oscillator
strength. The resulting absorption line, which can be called the {\it
pinning mode\/}, was first discussed by Fukuyama and Lee
(FL).~\cite{Fukuyama_77,Fukuyama_78} In those early works the linewidth
$\Delta\omega_p$ of the pinning mode was predicted to be of the order of
$\omega_p$ itself. Strictly speaking, FL considered not a WC but a more
conventional charge-density wave where the charge modulation has only
one harmonic, i.e., is cosine-like. Experiments on such materials, e.g.,
${\rm K}_{0.3}{\rm Mo}{\rm 0}_3$ (``blue bronze''), ${\rm Ta} {\rm
S}_3$, {\it etc.\/}, performed in {\it zero magnetic field\/} indeed
revealed broad absorption lines at disorder-dependent
frequencies.~\cite{Degiorgi_91} A later work of Normand {\it
et~al.\/}~\cite{Normand_92} devoted specifically to the WC in a strong $B$
revised some results of FL but left unchanged the prediction that
$\Delta\omega_p \sim \omega_p$. At the time, experiments seemed to
confirm that.~\cite{Palaanen_92} It came as a surprise when most recent
measurements in strong magnetic fields~\cite{Li_97,Engel_97,Li_98,Beya}
demonstrated that $\Delta\omega_p$ can be more than order of magnitude
smaller than $\omega_p$. Such unexpected findings revived interest in
this long-standing problem.

An important step towards resolution of the puzzle has been made by
Fertig,~\cite{Fertig_99} who showed that long-range Coulomb interaction
plays an important role in reducing the inhomogeneous broadening of the
pinning mode. However, Fertig used an oversimplified model (commensurate
pinning) and did not calculate $\Delta\omega_p$ directly. His results
for another quantity may be interpreted as an indication that
$\Delta\omega_p / \omega_p$ is exponentially small for weak pinning.
More realistic model was studied by Chitra~{\it
et~al.\/}~\cite{Chitra_98} who considered pinning by a short-range
Gaussian random potential of a general type. We comment on their results
later in this section.

In the present paper we study essentially the same model as FL and
Chitra~{\it et~al.\/} except we treat the electrons classically. This
limits the applicability of our results to the case $\xi > l_B$, where
$\xi$ is the correlation length of the pinning potential and $l_B =
\sqrt{\hbar c / e B}$ is the magnetic length. The classical
approximation enables us to focus on the interaction of the WC with
disorder, which is really the essence of the pinning mode
phenomenon. We critically re-examine the previous work on the subject,
identify the physical mechanism responsible for the line narrowing,
and spell out the conditions under which it occurs:
\begin{itemize}
\item the lattice dynamics is inertial (not overdamped)

\item magnetic field is sufficiently strong, such that
$\omega_c$ is larger than $\omega_p$

\item the pinning is sufficiently weak so that $\omega_p$
is much smaller than the magnetophonon bandwidth

\item the compression modulus $\lambda$ of the WC
evaluated at the pinning length is larger than its shear modulus
$\mu$ (satisfied for long-range interaction).

\end{itemize}
If any of the conditions above is violated, then the
line is conventionally broad, $\Delta\omega_p \sim \omega_p$.

Although we were unable to find the exact expression for
$\Delta\omega_p$, we show that there is an asymptotically exact
relation between $\Delta\omega_p$ at intermediate $B$ and the
low-frequency tail of the phonon spectral function in {\it zero\/} $B$
(we remind the reader here that we study classical electrons, which
form the WC at any $B$). Establishing such a connection and elucidating
the physical mechanism of the line narrowing are our main achievements.

\begin{figure}
\centerline{
\psfig{file=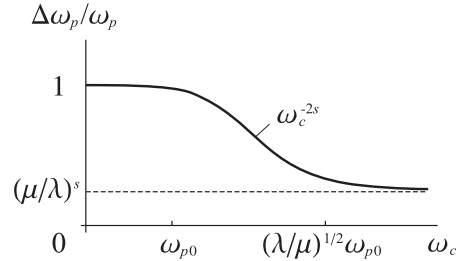,width=2.3in,%
bbllx=66pt,bblly=431pt,bburx=422pt,bbury=641pt}}
\vspace{0.1in}
\setlength{\columnwidth}{3.2in}
\centerline{\caption{Relative linewidth of the pinning mode as a function
of the cyclotron frequency.
\label{Fig_linewidth}
}}
\end{figure}

The phonon spectral function at low frequencies is expected to have a
power law form
\begin{equation}
  {\rm Im} D_{\alpha\alpha}(q = 0, \omega) \propto \omega^{2 s + 1},
\label{Im_D_omega}
\end{equation}
with a nontrivial exponent $s$. This additional input and further
considerations enable us to predict that as $\omega_c$ increases and
becomes larger than $\omega_p$, $\Delta\omega_p / \omega_p$ first
decreases in a power-law fashion 
\begin{equation}
 \Delta\omega_p / \omega_p \sim (\omega_p / \omega_c)^s
\label{linewidth}
\end{equation}
but then eventually saturates at a value (see
Fig.~\ref{Fig_linewidth})
\begin{equation}
 \Delta\omega_p / \omega_p \sim (\mu / \lambda)^s
\label{linewidth_strong}
\end{equation}
(here $\lambda$ is meant to be evaluated at the pinning length). The recent
experiments~\cite{Li_97,Engel_97,Li_98,Beya} have probed the high-field
regime where Eq.~(\ref{linewidth_strong}) is supposed to apply.

In the aforementioned work of Chitra~{\it et~al.\/},~\cite{Chitra_98} a
set of equations was derived, which, when analyzed further, also yield
Eq.~(\ref{linewidth_strong}), with $s = \frac12$. This specific value of
$s$ can be traced down to the fact that Chitra~{\it et~al.\/}
unknowingly rederive a common form of the self-consistent Born
approximation (SCBA), where Eq.~(\ref{Im_D_omega}) with $s = \frac12$ is
satisfied. Since the SCBA is uncontrolled at small $\omega$, these
results are unreliable. Moreover, appealing to the known properties of
Lifshitz tails~\cite{Lifshitz_book} in other disordered systems, we
argue that the SCBA strongly overestimates the low-frequency spectral
function. We propose that the correct approach should be based on
considering certain low-propability disorder configurations (``optimal
fluctuations''), which leads to
\begin{equation}
                                 s = \frac32.
\end{equation}
%

%
%

The paper is organized as follows. In Sec.~\ref{Statics} we discuss
the ground state of the pinned WC, placing emphasis on the parameters
that provide the input for the dynamical problem of interest. In
Sec.~\ref{Model} we formulate the model and the general framework for
study of the finite-frequency response. In Sec.~\ref{L_scattering} we
derive Eq.~(\ref{linewidth}). In Sec.~\ref{Soft_modes} we study the
soft phonon modes at $B = 0$ and obtain the above formula for
$s$. The comparison with the experiments is given in
Sec.~\ref{Discussion} after a brief discussion of quantum and
finite-temperature effects in Sec.~\ref{Quantum_Thermal}.

\section{Static properties}
\label{Statics}

The distortions of the lattice in the ground state of a collectively
pinned Wigner crystal accumulate gradually over length scales much
larger than the lattice constant $a$. Such distortions can be described
in terms of a smooth displacement field ${\bf u}^{(0)}({\bf r})$. The
Hamiltonian of the system $H = H_{el} + H_p$ is the sum of
the elastic term, 
\begin{eqnarray}
&\displaystyle H_{el} = \frac12 \int\limits_{\bf q}
u_\alpha^{(0)}(-{\bf q}) H_{\alpha\beta}^0({\bf q})
u_\beta^{(0)}({\bf q}),&
\label{H_el}\\
&\displaystyle H^0_{\alpha\beta} = 
\left(\delta_{\alpha\beta} - \frac{q_\alpha q_\beta}{q^2}\right)
H^0_{T}(q) + \frac{q_\alpha q_\beta}{q^2} H^0_{L}(q),&
\label{H_T}\\
&\displaystyle H^0_{T}(q) = \mu q^2, \:\:\: H^0_{L}(q) = \lambda(q) q^2,&
\label{H_L}
\end{eqnarray}
and the pinning term,~\cite{Giamarchi_95}
\begin{eqnarray}
  &\displaystyle H_p = n_e \!\! \int\limits_{\bf r} U({\bf r})
   \left\{\sum_{\bf K} \cos {\bf K} [{\bf r}- {\bf u}^{(0)}({\bf r})]
     - \bbox{\nabla}{\bf u}^{(0)}\right\},&
\label{H_p}\\
  &\displaystyle\langle U({\bf r}) U(0) \rangle = C(r).&
\label{C_r}
\end{eqnarray}
The notations here are as follows. The ${\bf q}$-integral is
taken over the Brillouin zone with measure $d^2 q / (2 \pi)^2$; the
${\bf r}$-integrals is taken over the area of the system with measure
$d^2 r$. Parameters $\mu$ and $\lambda$ represent the shear and bulk
elastic moduli of the WC, respectively. In strong magnetic fields, i.e.,
at low filling factors they approach their classical
values,~\cite{Bonsall_77} $\mu \approx 0.245 e^2 n_e^{3/2} / \kappa$
and $\lambda(q) = -5 \mu + Y / |q|$. Here $n_e = 2 / \sqrt{3}\, a^2$ is
the average electron concentration, $\kappa$ is the dielectric constant
of the medium, and $Y = 2 \pi e^2 n_e^2 / \kappa$. The infrared
divergence of $\lambda(q)$ originates from the long-range Coulomb
interaction. ${\bf K}$'s are the reciprocal lattice vectors. Finally,
the correlator $C(r)$ is assumed to rapidly decay at $r > \xi$ where $\xi
< a$ is the correlation length of the short-range Gaussian random
potential $U({\bf r})$.

The static properties of the system defined by
Eqs.~(\ref{H_el}--\ref{C_r}) can be investigated using classical
statistical mechanics methods. Problems of this type have been studied
extensively in the past, see Ref.~\onlinecite{Blatter_94} for review
and references, and also some recent contributions,
Refs.~\onlinecite{Giamarchi_95},
\onlinecite{Natterman_95}--\onlinecite{Giamarchi_xxx}. The
2D WC generally conforms to the $d = 2$, $m = 2$ class of elastic
media ($d$ is the total number of spacial dimensions and $m$ is the
number of components of ${\bf u}$). However, some caution is required
in transferring the known results to the WC case because the L- and
T-components are not equivalent. The large disparity $\lambda \gg \mu$
of the elastic moduli causes strongly suppression of the longitudinal
distortions compared to the transverse ones. Therefore, in certain
formulas one has to use $m = 1$. In this respect, the WC is similar to
the Abrikosov vortex lattice where the analogous inequality $c_{11}
\gg c_{66}$ exists.~\cite{Blatter_94}

\begin{figure}
\centerline{
\psfig{file=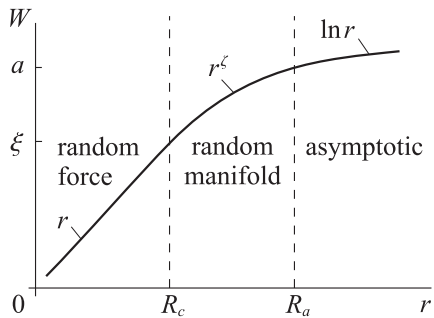,width=2.2in,%
bbllx=100pt,bblly=433pt,bburx=477pt,bbury=709pt}}
\vspace{0.1in}
\setlength{\columnwidth}{3.2in}
\centerline{\caption{Roughness $W(r)$.
\label{Fig_W}
}}
\end{figure}

Let us define the two-point correlation function
(roughness) $W(r) = \langle [{\bf u}^{(0)}({\bf r}) - {\bf
u}^{(0)}(0)]^2 \rangle^{1/2}$. Many static properties of the system can
be deduced from the behavior of $W(r)$. Three regimes can be
distinguished: random-force, random-manifold, and the asympotic
(Fig.~\ref{Fig_W}).

%
%

The {\it random-force\/} or Larkin regime appears for small $r$ where
$W(r) < \xi$. The last inequality legitimizes expanding the exponential in
Eq.~(\ref{H_p}) and keeping only the linear in $u^{(0)}$ terms. The
Hamiltonian becomes quadratic, and its ground state is easily found to
be
\begin{equation}
u_\alpha^{(0)}({\bf q}) =
-i n_e \sum_{{\bf K}} [{\bf H}^0]^{-1}_{\alpha\beta}({\bf q})
K_\beta \tilde{U}({\bf q} + {\bf K}).
\label{u_0_expanded}
\end{equation}
Here and below tildes denote Fourier transforms.

Averaging over the disorder, we arrive at 
\begin{eqnarray}
\displaystyle W^2({\bf r}) &\simeq& V^\prime(0) \!\!
\int\limits_{\bf q}
(1 - e^{i {\bf q}{\bf r}})\, \left[(H^0_T)^{-2}
+ (H^0_L)^{-2}\right]
\label{B_Larkin}\\
\displaystyle  &\propto& r^2 \ln (R_c / r),
\label{W_Larkin}\\
\displaystyle V(z) &\equiv& -\frac{n_e^2}{2} \sum_{{\bf K}}
\tilde{C}(K) e^{-z K^2 / 2}.
\label{V}
\end{eqnarray}
To obtain Eq.~(\ref{W_Larkin}) we cut off the log-divergent integral in
Eq.~(\ref{B_Larkin}) at $q = q_c$ where $q_c = \pi / R_c$ and $R_c$ is
the Larkin length defined by the relation $W(R_c) = \xi$. For a weak
random potential $R_c$ is much larger than $a$ and is given by 
\begin{equation}
                      R_c \sim \mu a \xi^2 [C(0)]^{-1/2}.
\label{R_c}
\end{equation}
The same estimate for $R_c$ can be obtained in a simpler
way,~\cite{Blatter_94,Fukuyama_78} by equating the elastic energy
density $\mu \xi^2 / R_c^2$ to the pinning energy density $[C(0)]^{1/2}
\sqrt{N} / R_c^2$. In the last expression $N = n_e R_c^2$ is the
number of electrons within the area $R_c \times R_c$ and the square root
takes into account that individual pinning energies have random signs.

The {\it elastic-manifold\/} regime is realized at intermediate
distances, $R_c < r < R_a$ where $\xi < W(r) < a$. Here the roughness
grows slower than in the Larkin regime, $W(r) \propto r^{\zeta}$, $\zeta
< 1$. For the 2D Wigner crystal the Flory estimate $\zeta = \frac13$ (as
appropriate for the $d = 2$, $m = 2$ case) should be accurate but
perhaps not exact.~\cite{Blatter_94} The slower growth rate of $W(r)$ is
due to a less efficient mechanism by which the lattice adjusts to the
random potential on length scales exceeding $R_c$. The steepest descent
towards the nearest local minima of the potential landscape in the
Larkin regime gets replaced by selection among a set of many possible
yet roughly equivalent nearby minima.

The third and the last regime begins when $W(r)$ reaches $a$ at the
length scale $R_a = R_c (a / \xi)^{1 / \zeta}$. The accumulation of the
pinning energy density at such large length scales becomes additionally
suppressed by lattice periodicity effects. Indeed, a uniform shift $u
\to u + a$ reduces to relabeling the lattice sites with no physical
significance. As a result, the roughness grows only logarithmically with
distance, $W(r) \sim a \ln(r / R_a)$. The asymptotic $\ln r$-roughness
corresponds to a ``quasi-short-range'' order: $W(r)$ increases so slowly
that it is barely enough to remove the algebraic divergencies (Bragg
peaks) of the static structure factor
\begin{equation}
 F({\bf q}) = n_e^2 \int\limits_{\bf r}
\exp\left[i ({\bf q} - {\bf K}){\bf r} - \frac12 K^2 W^2(r)\right]
\label{Structure_factor}
\end{equation}
at the reciprocal lattice vectors ${\bf q} = {\bf K}$. Remarkably,
it also creates enough elastic stress to generate unbound dislocations,
on the length scale~\cite{Giamarchi_xxx} $\xi_D \sim R_a \exp [C_0
\ln^{1/2} (R_a / a)]$. At this scale ${\bf u}$ becomes multi-valued and
Eq.~(\ref{H_el}) must be modified.

Although very interesting, the random-manifold and the asymptotic
regimes do not play much role in the further discussion. We mentioned
them here for the sake of a more comprehensive introduction and also
to draw attention to the difference between two characteristic length
scales, $R_c$ and $R_a$. $R_a$ is the length that seems to be
important only for rather subtle properties mentioned above: the
destruction of the Bragg peaks and the appearance of the topological
defects. In contrast, $R_c$ determines most of the physically
important response characteristics such as the threshold electric
field in the dc transport and the frequency $\omega_p$ of the pinning
mode.~\cite{Chitra_98} This role of $R_c$ originates from the fact
that it is the length scale where the dominant contribution $\sim \mu
\xi^2 / R_c^2$ to the pinning energy density is accumulated. In this
sense $R_c$ plays the role of the pinning length mentioned in
Sec.~\ref{Introduction}.

Let us now elaborate on the difference between the T- and L-correlation
functions mentioned above. To do so we decompose the elastic
displacement into its T- and L-components,
\begin{equation}
     {\bf u}({\bf q}) = \hat{\bf q} u_L({\bf q})
        + [\hat{\bf z} \times \hat{\bf q}] u_T({\bf q}),
\label{LT_decomposition}
\end{equation}
where $\hat{\bf q} = {\bf q} / |{\bf q}|$, and define the two components
of roughness $W_\alpha = \langle u_\alpha^{(0)}({\bf r}) -
u_\alpha^{(0)}(0)\rangle^{1/2}$, $\alpha = L, T$. First of all, let us
examine the short-distance behavior using Eq.~(\ref{u_0_expanded}). For
$W_L(r)$ we obtain the expression similar to Eq.~(\ref{B_Larkin}):
\begin{equation}
 W^2_L(r) \simeq V^\prime(0) \!\!
\int\limits_{\bf q}
\frac{1 - e^{i {\bf q}{\bf r}}}{[H^0_L(q)]^2}
\simeq \frac{V^\prime(0)}{2 \pi Y^2}
\ln\left(\frac{r}{a}\right),
\label{W2_L}
\end{equation}
and so $W_L(r)$ is much smaller than $W_T(r) \simeq W(r)$ given by
Eq.~(\ref{W_Larkin}). Formula~(\ref{W2_L}) is certainly valid for $r <
R_c$. For larger $r$ we can no longer treat the elastic distortion
$|{\bf u}^{(0)}({\bf r}) - {\bf u}^{(0)}(0)|$ as small. Nonetheless, we
expect that the region of validity of Eq.~(\ref{W2_L}) extends somewhat
beyond $r = R_c$. Indeed, instead of doing the perturbation theory
around the ideal lattice state, we can do it around the ground state
${\bf u}^{(0)}_\infty$ of the incompressible crystal, $\lambda =
\infty$. In the second approach we have to account for a finite pinning
energy density, but as long as the longitudinal elastic stiffness $Y q$
where $q \sim \pi / r$ exceeds the characteristic curvature $\mu q_c^2$ of
the pinning energy density landscape, i.e., for $r \lesssim R_L = (Y /
\mu) R_c^2 \sim R_c^2 / a$, this should be unimportant. Substituting $r
= R_L$ into Eq.~(\ref{W2_L}) we find that $W_L(R_L) \ll \xi$, contrary to
some previous suggestions~\cite{Normand_92} that $W_L(R_L) \sim
W_T(R_L)$ [recall that $W_T(R_L) \simeq W(R_L) \gg W(R_c) = \xi$].

The large-distance ($r > R_L$) behavior of $W_L(r)$ is unaccessible
within the perturbation theory. However, we should get qualitatively
correct results by using the Gaussian variational replica method
(GVRM).~\cite{Giamarchi_95} Assuming for simplicity that $R_L \gg R_a$
and suitably modifying GVRM for the present problem, we find
\begin{equation}
     W_L^2(r) \propto V^\prime(0) \int\limits_{\bf q}
\frac{1 - \cos({\bf q}{\bf r})}
     {H^0_L(q) [H^0_L(q) + \Pi_a]},
\label{W_L_GVM}
\end{equation}
where $\Pi_a = \mu R_a^{-2}$. It is easy to see that $W_L(r)$ tends
to a finite limiting value as $r \to \infty$. So, the long-range positional
order is absent in the T- but is preserved in the L-part of the
elastic displacement field.~\cite{Comment_on_W_and_B}

To conclude the review of the statics, let us consider a matrix
of the second derivatives of $H_p$,
\begin{equation}
{\bf S}({\bf r}) = \frac{\partial^2 H_p}
             {\partial {\bf u}^{(0)} \partial {\bf u}^{(0)}}
                 = n_e \sum_{{\bf K}} {\bf K}
{\bf K} U({\bf r}) \cos {\bf K} [{\bf r}- {\bf u}^{(0)}({\bf r})].
\label{S}
\end{equation}
Matrix ${\bf S}$ plays an important role in the dynamics and we need to
understand its properties.

In the case of a short-range disorder we study here, the correlations in
$U({\bf r})$ decay more rapidly with $r$ than those of ${\bf
u}^{(0)}({\bf r})$. This enables us to find the two-point correlator of
$S_{\alpha\beta}$ with little effort. It suffices to use the
approximation ${\bf u}^{(0)}({\bf r}) = 0$ in Eq.~(\ref{S}), which leads
to
\begin{eqnarray}
&\displaystyle \langle \tilde{S}_{\alpha\mu}({\bf k}_1)
 \tilde{S}_{\nu\beta}({\bf k}_2) \rangle
- \langle \tilde{S}_{\alpha\mu} \rangle
\langle \tilde{S}_{\nu\beta} \rangle
 \simeq -\frac{V^{\prime\prime}(0)}{2}&\nonumber\\
&\displaystyle \times (\delta_{\alpha\mu} \delta_{\nu\beta}
+ \delta_{\alpha\beta} \delta_{\mu\nu}
+ \delta_{\alpha\nu} \delta_{\mu\beta}) \delta_{{\bf k}_1 {\bf k}_2} L^2&
\label{S_rms}
\end{eqnarray}
for the WC with the hexagonal lattice and in the long-wavelength
limit $k_1, k_2 \ll 1 / a$. Here $L^2$ is the area of the system.
Another important property of $S_{\alpha\beta}$ is a nonzero mean,
\begin{equation}
\langle S_{\alpha\beta}({\bf r})\rangle =
\delta_{\alpha\beta} S_0.
\label{S_0}
\end{equation}
Parameter $S_0$ is positive, in accordance with the pinning phenomenon:
the crystal in its ground state is distorted in such a way
that the electrons are preferentially situated near the minima of
the random potential $U({\bf r})$ where its curvature is positive. $S_0$
can be estimated to be
\begin{equation}
S_0 \simeq -\frac{V^{\prime\prime}(0)}{2\pi\mu}
\ln\left(\frac{R_c}{a}\right)
\label{S_0_Larkin}
\end{equation}
using the same procedure as for deriving Eq.~(\ref{u_0_expanded}).
Assuming that $|{\bf u}^{(0)}({\bf r}) - {\bf u}^{(0)}(0)|$ is
``small,'' we expand the exponential in Eq.~(\ref{S}) to obtain
\begin{equation}
\tilde{S}_{\alpha\beta}({\bf q}) \simeq
-\frac{i n_e}{L^2} \sum_{{\bf K}} K_\alpha
K_\beta K_\gamma \int\limits_{{\bf q}_1}
u_\gamma^{(0)}({\bf q}_1)
\tilde{U}({\bf q} - {\bf q}_1 - {\bf K}).
\label{S_0_expanded}
\end{equation}
Combining Eqs.~(\ref{u_0_expanded}) and (\ref{S_0_expanded}), averaging
over the disorder, and using the inequality $H^0_L(q) \gg H^0_T(q)$, we
get
\begin{equation}
S_0 \simeq -V^{\prime\prime}(0) \int\limits_{\bf q} [H^0_T(q)]^{-1},
\label{S_0_Larkin_II}
\end{equation}
which leads to Eq.~(\ref{S_0_Larkin}). Here we again had to cut off the
infrared logarithmic divergence by hand, at $q = q_c = \pi / R_c$. More
sophisticated approximation schemes such as the GVRM would implement this
cutoff more gracefully but as long as we are not interested in the
numerical factor in the argument of the log, the result is the same.

At this point we conclude the discussion of the ground state
properties of the pinned WC, as we are now ready to address its dynamical
response.

\section{Finite-frequency response: perturbation theory
and qualitative considerations}
\label{Model}

To study the dynamics of the WC we introduce a time-dependent
displacement field ${\bf u}$, which is the deviation of the total
displacement from its ground-state value ${\bf u}^{(0)}({\bf r})$. We
will restrict ourselves to the harmonic approximation where the action
is quadratic in ${\bf u}$. The response of a harmonic oscillator is the
same in quantum and classical mechanics, so we can use the convenient
imaginary-time formalism without compromising our original
intention to treat the system classically. The resultant action contains
two terms: one which describes a uniformly pinned WC,
\begin{equation}
A_0 = \frac12 \frac{1}{\hbar\beta} \sum\limits_{\omega_n}
\int\limits_{\bf q}
{\bf u}^\dagger({\bf q}, i \omega_n) [{\bf D}^0]^{-1}
{\bf u}({\bf q}, i \omega_n),
\label{A_0}
\end{equation}
and the other which takes into account fluctuations in local pinning
strength,
\begin{equation}
A_1 = \frac12 \frac{1}{\hbar\beta} \sum\limits_{\omega_n}
                 \int\limits_{{\bf q}_1} \int\limits_{{\bf q}_2}
{\bf u}^\dagger({\bf q}_1, i \omega_n)
\delta\tilde{\bf S}({\bf q}_1 - {\bf q}_2)
{\bf u}({\bf q}_2, i \omega_n).
\label{A_1}
\end{equation}
Here $\beta = 1 / (k_B T)$ is the inverse temperature, $\omega_n = 2 \pi n
/ (\hbar\beta)$ are the bosonic Matsubara frequencies, $\delta\tilde{\bf
S} \equiv \tilde{\bf S} - S_0 {\bf I}$, ${\bf I}$ is the identity
matrix, ${\bf D}^0 = {\bf D}^0({\bf q}, i \omega_n)$ is the phonon
propagator of a uniformly pinned WC (cf.~Refs.~\onlinecite{Fukuyama_78}
and \onlinecite{Bonsall_77})
\begin{equation}
     [{\bf D}^0]^{-1} = {\bf R}^\dagger_{\bf q} \left[
\begin{array}{cc}
H^0_T + S_0 + \rho \omega_n^2     & -\rho \omega_n \omega_c\\
\rho \omega_n \omega_c            & H^0_L + S_0 + \rho \omega_n^2
\end{array}\right]
{\bf R}_{\bf q},
\label{D_0}
\end{equation}
$\rho = m_e n_e$ is the average mass density, and ${\bf R}_{\bf q}$ is the
$O(2)$ rotation by angle $\arg(q_x + i q_y)$. We also define the
disorder-averaged propagator ${\bf D}({\bf q}, i \omega_n)$,
\begin{equation}
{\bf D}({\bf q}, i \omega_n) = 
\left\langle \frac{\int {\cal D}{\bf u} {\cal D}{\bf u}^*
{\bf u}({\bf q}, i \omega_n) {\bf u}^*({\bf q}, i \omega_n) e^{-A}}
{(\hbar \beta L^2) \int {\cal D}{\bf u} {\cal D}{\bf u}^* e^{-A}}
\right\rangle,
\label{D_def}
\end{equation}
where $A \equiv A_0 + A_1$ and $L^2$ is again the area of the system.

The quantity we set out to calculate is the ac conductivity
$\bbox{\sigma}(q, \omega)$, which in this model is
given by~\cite{Fukuyama_78}
\begin{equation}
\bbox{\sigma}(q, \omega) = -i e^2 n_e^2 \omega
\left.{\bf D}(q, i \omega_n)\right|_{i \omega_n \to \omega + i 0}.
\label{sigma}
\end{equation}
The conductivity can also be expressed in terms of the phonon
self-energy
\begin{equation}
{\bf \Pi} ({\bf q}, \omega) \equiv S_0 {\bf I} +
\left.\left({\bf D}^{-1} - [{\bf D}^0]^{-1}\right)
\right|_{i \omega_n \to \omega + i 0}.
\label{Pi_def}
\end{equation}
We are interested primarily in the case $q = 0$ where the most general
form of ${\bf \Pi}({\bf q}, \omega)$ consistent with rotational
symmetry is
\begin{equation}
\Pi_{\alpha\beta}(0, \omega) = \delta_{\alpha\beta} \Pi(\omega)
- i \varepsilon_{\alpha\beta} \rho \omega \omega_c f_{x y}(\omega),
\label{Pi_q=0}
\end{equation}
$f_{x y}(\omega)$ being the relative correction $\Delta \rho_{x y} /
\rho^0_{x y}$ to the bare Hall resistivity $\rho^0_{x y}(\omega) = B /
(n_e e c)$. Approximate calculations presented in this paper give $f_{x
y} = 0$, which leads us to believe that $f_{x y}(\omega)$ must be small
for weak pinning. We choose to neglect such fine details and to assume
that $f_{x y}$ vanishes.~\cite{Comment_on_Hall} In this case
$\bbox{\Pi}(0, \omega)$ is a scalar and $\sigma_{xx}(\omega)$ is given
by
\begin{equation}
{\rm Re}\,\sigma_{xx}(\omega) = e^2 n_e^2 \omega\, {\rm Im}\,
\frac {\Pi(0, \epsilon) - \epsilon}
{[\Pi(0, \epsilon) - \epsilon]^2 - \epsilon \epsilon_c},
\label{sigma_xx}
\end{equation}
where we introduced convenient ``energy'' variables
\begin{equation}
 \epsilon \equiv \rho \omega^2, \quad
 \epsilon_c \equiv \rho \omega_c^2.
\label{epsilon_def}
\end{equation}
From Eq.~(\ref{sigma_xx}) one can see that in strong magnetic fields,
$\omega_c \gg \omega_p$, the power absorption in the uniform electric
field takes place mainly within the frequency interval of width
\begin{equation}
\Delta\omega_p = -\frac{{\rm Im}\,\Pi(0, \epsilon_p)}{\rho \omega_c}
\label{Delta_omega_p_def}
\end{equation}
centered at $\omega = \omega_p$, where
\begin{equation}
\omega_p = \sqrt{\frac{\epsilon_p}{\rho}}
= \frac{{\rm Re}\, \Pi(0, \epsilon_p)}{\rho \omega_c}.
\label{omega_p_def}
\end{equation}
The last equation is the implicit definition of $\epsilon_p$. In the
following we will work predominantly with the ``energy'' variable
$\epsilon$ rather than the frequency $\omega$.

At small $\epsilon$ the calculation of $\bbox{\Pi}$ is a difficult
problem but at large energies, $\epsilon \gg \epsilon_p$, the
first Born approximation suffices. The only parameters
needed to implement it are the mean and the variance
of the matrix elements of ${\bf S}$ given by Eqs.~(\ref{S_rms}) and
(\ref{S_0}). For $q \ll 1 / a$ we obtain that $\Pi_{\alpha\beta}({\bf q},
\epsilon) = \delta_{\alpha\beta} \Pi(\epsilon)$, where
\begin{equation}
\Pi(\epsilon) =  S_0 + V^{\prime\prime}(0)
\int\limits_{\bf k} {\rm tr}\,{\bf D}^0({\bf k}, \epsilon).
\label{Pi_Born}
\end{equation}
$\Pi$ has both imaginary and real parts, corresponding to the broadening
and the frequency shift of the pinning mode,
Eqs.~(\ref{Delta_omega_p_def}) and (\ref{omega_p_def}), respectively.
The imaginary part comes solely from the pole(s) of ${\bf D}^0({\bf k},
\epsilon)$, i.e., the solutions of $[H^0_T({\bf k}) + S_0 - \epsilon]
[H^0_L({\bf k}) + S_0 - \epsilon] - \epsilon \epsilon_c = 0$. A more
accurate expression for $\Pi$ can be obtained using the {\it
self-consistent\/} Born approximation (SCBA),
\begin{equation}
\Pi(\epsilon) =  S_0 + V^{\prime\prime}(0)
\int\limits_{\bf k} {\rm tr}\,{\bf D}({\bf k}, \epsilon),
\label{Pi_SCBA}
\end{equation}
as long as the full propagator ${\bf D}({\bf k}, \epsilon)$ has a pole
at $|k| \gg q_c = \pi / R_c$. Under this condition the diagrams with
intersecting lines not included into the SCBA are suppressed, much like
they are suppressed in a dirty metal with $k_F \gg l^{-1}$, $k_F$ being
the Fermi momentum and $l$ being the mean free path. The indicated
condition is satisfied when
\begin{equation}
\epsilon \gg \min
\left\{\Pi_0, \frac{\lambda}{\mu} \frac{\Pi_0^2}{\epsilon_c}\right\}.
\label{SCBA_domain}
\end{equation}
Here and below $\lambda$ is meant to be evaluated at $q = q_c$ unless it
is indicated otherwise. Parameter $\Pi_0$, which can be estimated to be
\begin{equation}
                            \Pi_0 \sim \mu q_c^{2}
\label{Pi_0_estimate}
\end{equation}
represents the real part of the self-energy at the lowest $\epsilon$
allowed by inequality~(\ref{SCBA_domain}). $\Pi_0$ is smaller than
$S_0$ by the logarithmic factor due to the partial cancellation
between the first and the second terms in Eq.~(\ref{Pi_SCBA}). This
unfortunate cancellation leaves us only with the order of magnitude
estimate~(\ref{Pi_0_estimate}).

%
%
\begin{figure}
\centerline{
\psfig{file=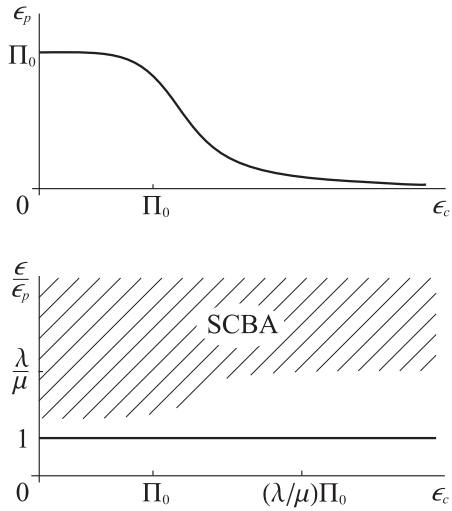,width=2.3in,%
bbllx=96pt,bblly=180pt,bburx=480pt,bbury=618pt}}
\vspace{0.1in}
\setlength{\columnwidth}{3.2in}
\centerline{\caption{Top: Pinning mode location as a function of the magnetic field.
Bottom: SCBA domain of applicability.
\label{Fig_SCBA_domain}
}}
\end{figure}

When inequality~(\ref{SCBA_domain}) is violated, the SCBA becomes an
uncontrolled approximation. The only SCBA result, which can presumably
be trusted at such $\epsilon$, is a slow dependence of ${\rm Re}\,
\Pi(\epsilon)$ on energy. This is because the real part of the integral
in Eq.~(\ref{Pi_SCBA}) is dominated by large $k$'s where the
lowest-order perturbation theory is valid. To estimate $\epsilon_p$ we
can take ${\rm Re}\, \Pi(\epsilon) \approx \Pi_0$, which
yields~\cite{Fukuyama_78,Normand_92}
\begin{equation}
\epsilon_p \sim \min \left\{\Pi_0, \frac{\Pi_0^2}{\epsilon_c}\right\}.
\label{epsilon_p_estimate}
\end{equation}
represented graphically in Fig.~\ref{Fig_SCBA_domain} together
with the SCBA domain.

\begin{figure}
\centerline{
\psfig{file=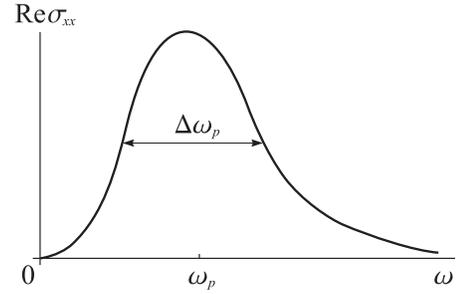,width=2.3in,%
bbllx=101pt,bblly=431pt,bburx=477pt,bbury=679pt}}
\vspace{0.1in}
\setlength{\columnwidth}{3.2in}
\centerline{\caption{Frequency dependence of the conductivity in
weak magnetic fields.
\label{Fig_sigma_weak_field}
}}
\end{figure}

As one can see from this graph, for $\epsilon_c < \Pi_0$ (weak magnetic
fields) there is no parametric separation between the SCBA domain and
the curve $\epsilon = \epsilon_p$; therefore, the SCBA is qualitatively
correct even along this curve where it predicts ${\rm Im}
\Pi(\epsilon_p) \sim -\Pi_0$. Hence, $\Delta\omega_p \sim \omega_p$. The
resultant broad absorption maximum is sketched
in Fig.~\ref{Fig_sigma_weak_field}. 

%
%

If $\lambda$ and $\mu$ are comparable, the same argument works
also in strong magnetic fields ($\epsilon_c \gg \Pi_0$) where
$\sigma_{xx}(\omega)$ looks qualitatively similar, except the position
of the maximum depends on the magnetic field: $\omega_p = \omega_{p 0}^2
/ \omega_c$, where $\omega_{p 0}$ is the pinning frequency at $B = 0$.
On the other hand, if $\lambda \gg \mu$, as for the WC, the pinning mode
is situated in the interior of a parametrically wide range of $\epsilon$
where the SCBA fails. Thus, the calculation of $\Delta\omega_p$ requires
other methods. Later we will show that $\Delta\omega_p$ in strong
magnetic fields is related to the low-frequency tail of
$\sigma_{xx}(\omega)$ {\it in the absence\/} of the magnetic field.
Although calculating the functional form of such a tail is another
nontrivial problem (after all, it is beyond the SCBA!), this intriguing
relation is sufficient to establish that the absorption line narrows
down, in agreement with the experiments.~\cite{Li_97,Engel_97,Li_98,Beya}

We unfold our argument gradually over the remaining sections. In this
section we start implementing this task by clarifying why
$\sigma_{xx}(\omega)$ is virtually independent of
the longitudinal stiffness $\lambda$ provided $\lambda \gg \mu$
and $B = 0$.

Since action $A$ is quadratic in ${\bf u}_L$, the L-phonon degrees of
freedom can be easily integrated out, leading to the effective
Hamiltonian for the T-phonons,
\begin{equation}
 {\bf H}_T = {\bf H}_T^0 + {\bf S}_T
     - {\bf S}_{X +}^\dagger ({\bf H}_L^0 + {\bf S}_L - \epsilon {\bf I})^{-1}
       {\bf S}_{X +},
\label{H_eff_T}
\end{equation}
where ${\bf H}_T^0$, ${\bf H}_L^0$, and ${\bf S}_i$
should be understood as operators. The
first two are diagonal in the basis of plane waves, with matrix elements
$H_T^0(q)$ and $H_L^0(q)$, respectively; ${\bf S}_i$'s have both
diagonal and off-diagonal matrix elements:
\begin{mathletters}
\begin{eqnarray}
S_{L\,{\bf q} {\bf q}^\prime} &=&
\tilde{S}_{\alpha\beta}({\bf q} - {\bf q}^\prime)
\hat{q}_{\alpha} \hat{q}_{\beta}^\prime,
\label{S_L}\\
S_{T\,{\bf q} {\bf q}^\prime} &=& (\delta_{\alpha\beta}\,
{\rm tr} \tilde{\bf S} -
\tilde{S}_{\alpha\beta}) \hat{q}_{\alpha} \hat{q}_{\beta}^\prime,
\label{S_T}\\
S_{X\,{\bf q} {\bf q}^\prime} &=&
\tilde{S}_{\alpha\beta}({\bf q} - {\bf q}^\prime)
\epsilon_{\beta\gamma}\hat{q}_{\alpha} \hat{q}_{\gamma}^\prime.
\label{S_X}
\end{eqnarray}
\end{mathletters}
Finally, ${\bf S}_{X +} \equiv {\bf S}_X + \sqrt{\epsilon \epsilon_c}\, {\bf I}$.
Integrating out ${\bf u}_L$ is an exact algebraic transformation, which
preserves the spectrum of the collective modes: the resolvent ${\bf
D}_T(\epsilon)$ of the operator ${\bf H}_T - \epsilon {\bf I}$ has poles at the
same $\epsilon$ as the full propagator. In fact, ${\bf D}_T$, which can
be written in the form
\begin{equation}
 {\bf D}_T = ([{\bf D}_T^*]^{-1} - {\bf S}^*)^{-1},
\label{D_T_unaveraged}
\end{equation}
where
\begin{eqnarray}
&&  {\bf D}_T^* = ({\bf H}_T^0 + {\bf S}_T - \epsilon {\bf I})^{-1} 
\label{D_T_*}\\
&&  {\bf S}^* = {\bf S}_{X +}^\dagger
         ({\bf H}_L^0 + {\bf S}_L - \epsilon {\bf I})^{-1} {\bf S}_{X +},
\label{S_*}
\end{eqnarray}
is nothing else than the T-T component of the full phonon propagator
before the disorder averaging. The self-energy of the averaged
propagator at $q = 0$ is a scalar (see above); therefore, the T-T
component determines the entire propagator for this particular $q$.
 
Let us now show that in zero magnetic field the mixing between the T-
and L-phonons represented by ${\bf S}^*$ has little effect on ${\bf
D}_T$. Indeed, in the diagrammatic expansion of $\langle {\bf
D}_T(\epsilon) \rangle$ in powers of $\delta{\bf S}_T$ and ${\bf S}^*$, the
typical momentum transfer $k$ for $\epsilon \lesssim \Pi_0$ is of the
order of $q_c$. A single occurence of ${\bf S}^*$ contributes $\sim
k^2 |V^{\prime\prime}(0)| / \lambda k^2 \sim (\mu / \lambda) \Pi_0$ to
the self-energy, compared to $\sim \Pi_0$ from $\delta{\bf S}_T$. Diagrams
with multiple occurences of ${\bf S}^*$ are suppressed by even higher
powers of the small parameter $\mu / \lambda$. Thus, in the parameter
range relevant for the observation of the pinning mode
\begin{equation}
\langle {\bf D}_T \rangle({\bf q}, \epsilon) \simeq
 \langle {\bf D}_T^* \rangle({\bf q}, \epsilon) \equiv
\frac{1}{H^0_T(q) + \Pi^*({\bf q}, \epsilon) - \epsilon}.
\label{D_T_averaged}
\end{equation}
Note that ${\bf D}_T^*$ is determined only by the shear modulus $\mu$
and the disorder in the T-T channel $\delta{\bf S}_T$, while bulk
modulus $\lambda$ drops out. It follows from this discussion that for
$\lambda \gg \mu$ we can calculate the $q = 0$ response pretending
that the L-degree of freedom does not exist.

Let us indeed imagine that the L-phonons are forbidden and try to
investigate the nature of the T-phonon eigenstates that would compose the
pinning mode, i.e., the states that would respond to the uniform
electric field (still at $B = 0$). To do so we need to analyze the
solutions of the eigenvalue problem for the single-particle Hamiltonian
${\bf H}^* = {\bf H}_T^0 + {\bf S}_T$. In the uniformly pinned WC, ${\bf
S}_T = S_0 {\bf I}$ and the eigenstates are just the plane waves
labelled by momenta ${\bf k}$ of the Brillouin zone. In the actual
random system the eigenstates are wavepackets of the plane waves with
characteristic spread of momenta of the order of $L_T^{-1}$, where $L_T
\sim v_T /\,|{\rm Im}\, \Pi|$ is the T-phonon mean free path, $v_T =
\sqrt{\epsilon \mu}$ playing the role T-phonon ``velocity.'' Since $|{\rm
Im}\, \Pi(\epsilon_p)| \sim \epsilon_p \sim \mu q_c^{2}$, at $\epsilon \sim
\epsilon_p$ we have $L_T \sim R_c$. Hence, the eigenstates that respond
appreciably to a $q = 0$ external drive are wavepackets built from $0
< k < q_c$ plane waves. For such wavepackets the average $k$ is
of the order of the inverse mean free path and the Ioffe-Regel
criterion~\cite{Ioffe_60} maintains that these states are
localized. In other words, $R_c$ is not only the mean free path but
also the localization length of the states within the pinning mode.

\begin{figure}
\centerline{
\psfig{file=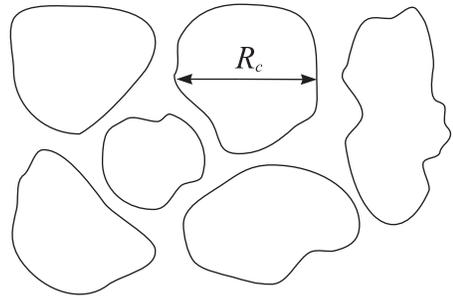,width=2.3in,%
bbllx=143pt,bblly=370pt,bburx=470pt,bbury=586pt}}
\vspace{0.1in}
\setlength{\columnwidth}{3.2in}
\centerline{\caption{Sketch of spacial distribution
of T-phonon states contributing to the
pinning mode at $B = 0$.
\label{Fig_localized_T}
}}
\end{figure}

Let us define the $B = 0$ T-phonon density of states,
\begin{equation}
\nu(\epsilon) = \frac{1}{L^2}
\left\langle \sum_i \delta(\epsilon - \epsilon_i)\right\rangle
= \frac{1}{\pi} \int\limits_{\bf k}
\langle {\bf D}^*_T \rangle ({\bf k}, \epsilon).
\label{nu_def}
\end{equation}
At $\epsilon \gg \epsilon_p$ it tends to the bare value $\nu_0 = (4 \pi
\mu)^{-1}$, and at $\epsilon \sim \epsilon_p$ it is only slightly
smaller. It is easy to see then that within an area $R_c \times R_c$ and
energy interval $0 < \epsilon < \epsilon_p$, there is typically only one
(localized) state. The picture that emerges is illustrated in
Fig.~\ref{Fig_localized_T}: we have a collection of localized states of
roughly the same size (i.e., the inverse participation ratio) and
roughly the same distance from each other in real space, both of the
order of $R_c$. The broad distribution of shapes and sizes
characteristic of localized states in disordered systems yields a large
inhomogeneous broadening $\Delta\omega_p \sim \omega_p$ of the
absorption line, as we found earlier based on a different line of
reasoning.

In early works on pinning a heuristic imagery of ``domains'' was
sometimes invoked~\cite{Normand_92} to describe the ground state
structure. However, it was never clear where to draw the boundaries
between different domains. This difficulty is partially resolved in our
picture of phonon localization where domains can be defined as areas
where individual localized phonon wavefunctions are appreaciable.
However, modern understanding of the subject reviewed in the previous
section no longer appeals to any ``domains,'' and so this interpretation
may be regarded as a historic sentiment.

%
%

One more historic comment is in order. Previous
work~\cite{Phonon_localization} on phonon localization found that the
phonon eigenstates remain extended even in the $\omega \to 0$ limit,
seemingly in contradiction to our results. In fact, there is no
contradiction because the model we study here and the conventional
formulations, where disorder originates from the defects of the
crystalline lattice, e.g., substitutions by impurity atoms, are quite
different. While in those conventional models the defects move with
the lattice and their influence rapidly diminishes at small ${\bf k}$'s,
the pinning potential for the WC is static, so it is not affected by
the lattice motion. It couples to the phonons via the matrix ${\bf S}$
whose fluctuations have approximately the same rms magnitude at all
${\bf k}$'s within the Brillouin zone, see Eq.~(\ref{S_rms}). This is
the reason why in our case the disorder effects are much stronger,
causing the localization of the phonon eigenstates. Actually, our
model has more in common with those of noninteracting electrons in
dirty metals and semiconductors. (At the one-particle level the
distinction between the bosonic statistics of phonons and the
fermionic statistics of electrons is irrelevant).

Concluding this section, we would like to reiterate one of its main
results, that the presence (or absence) of the stiff L-phonon degree of
freedom affects the $B = 0$, $q = 0$ response only weakly. Naturally, it
makes the fate of the L-phonons seem quite mysterious. The finite-$B$
dynamical response has also been barely touched upon. These gaps in
understanding will be filled in the next section.

\section{Scattering of L-phonons and pinning mode linewidth
in a finite magnetic field}
\label{L_scattering}

In the previous section we discussed the scattering and
localization of the T-component of WC lattice vibrations. Let us
now turn to the L-component. This time we integrate out ${\bf u}_T$
to obtain the effective Hamiltonian for the L-phonons,
the corresponding Green's function, and the self-energy:
\begin{eqnarray}
&& {\bf H}_L = {\bf H}_L^0 + {\bf S}_L
     - {\bf S}_{X +} ({\bf H}_T^0 + {\bf S}_T - \epsilon {\bf I})^{-1}
       {\bf S}_{X +}^\dagger,
\label{H_eff_L}\\
&& {\bf D}_L = ({\bf H}_L - \epsilon {\bf I})^{-1},
\label{D_L_def}\\
&& \bbox{\Pi}_L(\epsilon) = \langle {\bf D}_L \rangle^{-1}
                          - ({\bf H}_L^0 - \epsilon {\bf I}).
\label{Pi_L_def}
\end{eqnarray}
Since the self-energy $\bbox{\Pi}({\bf q}, \epsilon)$ of the original
system (before integrating out ${\bf u}_T$) is a scalar at ${\bf q} = 0$,
$\Pi_L$ must obey the relation
\begin{equation}
\Pi_L(0, \epsilon) = \Pi(0, \epsilon)
 - \frac{\epsilon \epsilon_c}{\Pi(0, \epsilon) - \epsilon}.
\label{Pi_L}
\end{equation}
Our first task is to reproduce the results of Sec.~\ref{Model} by
demonstrating that in sufficiently weak magnetic fields $\Pi_L$ indeed
has the above form with $\Pi \simeq \Pi^*$.

\begin{figure}
\centerline{
\psfig{file=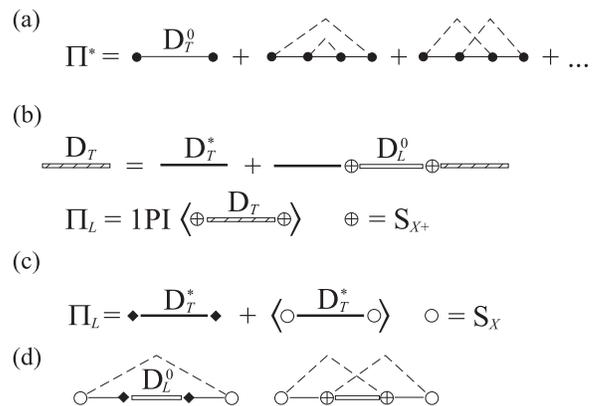,width=3.0in,%
bbllx=77pt,bblly=369pt,bburx=533pt,bbury=687pt}}
\vspace{0.1in}
\setlength{\columnwidth}{3.2in}
\centerline{\caption{(a) Diagrams for $\Pi^*$. Filled circles and solid
lines represent $\delta{\bf S}_T$ and ${\bf D}_T^0 \equiv ({\bf H}^0_T +
S_0 {\bf I} - \epsilon {\bf I})^{-1}$, respectively. Dashed lines
symbolize contractions. (b) Diagrammatic definitions of ${\bf D}_T$ and
$\Pi_L$. (c) Diagrams without ${\bf S}^*$. Diamonds represent factors
$i\protect\sqrt{\epsilon \epsilon_c}$.
(d) Some representive diagrams for $\Pi_L$ not included in (c). 
\label{Fig_Pi_L}
}}
\end{figure}

There are two types of processes that contribute to $\Pi_L$: the
intraband scattering (L-L channel) and the scattering through an
intermediate T-state (L-T-L channel). They originate from the second and
the third terms of ${\bf H}_L$, respectively. The intraband scattering
will be treated in more detail in Appendix~\ref{Supersymmetry} where
using a combination of the renormalization group and instanton methods
we show that it is exponentially small. For our current purposes a
weaker result is sufficient, that the intraband contribution $\Pi_{L \to
L}$ to $\Pi_L$ is much smaller than $\Pi_0$. This can be established
without detailed calculations. Indeed, the effective random potential
${\bf S}_L$ in the L-L channel has basically the same mean and rms
fluctiations as ${\bf S}_T$; therefore, ${\bf S}_L$ and ${\bf S}_T$ can
be regarded as disorder of approximately the same strength. Then the
alluded inequality $|\Pi_{L \to L}| \ll |\Pi| \sim \Pi_0$ follows simply
from the fact that L-phonons have much steeper bare dispersion relation
$\epsilon(q) = \lambda(q) q^2$ than T-phonons and are scattered much
weaker by disorder of the same strength. It is legitimate to neglect the
L-L scattering altogether by replacing ${\bf S}_L$ with its average
value $S_0 {\bf I}$, and to concentrate exclusively on the L-T-L
processes. In this case, much like for a dirty semiconductors with two
bands of carriers,~\cite{Ablyazov_91} the diagrammatic expansion of
$\Pi_L$ can be formulated as the sum of all one-particle irreducible
(1PI) graphs involving ${\bf S}_{X +}$ and ${\bf D}_T$, see
Eqs.~(\ref{D_T_unaveraged}--\ref{S_*}) and Fig.~\ref{Fig_Pi_L}a--b.

%
%

The crucial point substantiated below is that ${\bf S}^*$ can be
neglected, i.e., that ${\bf D}_T$ can be replaced by ${\bf D}^*_T$,
provided $\lambda \gg \mu$ and the magnetic field is not too strong,
\begin{equation}
         \epsilon_c \ll ({\lambda}/{\mu}) \Pi_0.
\label{Intermediate_field}
\end{equation}
For such $\epsilon_c$ only two graphs contribute to $\Pi_L$
(Fig.~\ref{Fig_Pi_L}c), which is a significant simplification. However,
we still have to specify how to average over the disorder. Naive
averaging of ${\bf S}_X$ and ${\bf D}^*_T$ separately from each other
yields
\begin{equation}
\Pi_L({\bf q}, \epsilon) =  S_0
- \epsilon \epsilon_c \langle {\bf D}^*_T \rangle({\bf q}, \epsilon)
+ V^{\prime\prime}(0)
\int\limits_{\bf k} \langle {\bf D}^*_T \rangle({\bf k}, \epsilon)
\label{Pi_L_naive}
\end{equation}
but it is certainly not accurate because ${\bf S}_X$ and $\delta{\bf
S}_T$ are not independent. Indeed, their two-point correlator
\begin{eqnarray}
\langle S_{X {\bf q}_1 {\bf q}} S_{T {\bf q}_2 {\bf q}_3}\rangle &\propto&
  (\hat{\bf q}_1 \hat{\bf q}_2) (\hat{\bf q}_3 \hat{\bf q})
+ (\hat{\bf q}_1 \hat{\bf q}_3) (\hat{\bf q}_2 \hat{\bf q})
\nonumber\\
&+& (\hat{\bf q}_1 \hat{\bf q}) (\hat{\bf q}_3 \hat{\bf q}_2)
\label{Correlator_S_X_S_T}
\end{eqnarray}
is in general nonvanishing.

In the special case of ${\bf q} = 0$ a further progress is possible.
Comparing the above equation with
\begin{eqnarray}
\langle S_{T {\bf q}_1 {\bf q}} S_{T {\bf q}_2 {\bf q}_3}\rangle &\propto&
  (\hat{\bf q}_1 \hat{\bf q}_2) [\hat{\bf q} \times \hat{\bf q}_3] \hat{\bf z}
+ (\hat{\bf q}_1 \hat{\bf q}_3) [\hat{\bf q} \times \hat{\bf q}_2] \hat{\bf z}
\nonumber\\
&+& (\hat{\bf q}_1 \hat{\bf q}) [\hat{\bf q}_3 \times \hat{\bf q}_2] \hat{\bf z},
\label{Correlator_S_T_S_T}
\end{eqnarray}
we observe that the former is transformed into the latter by the
replacement ${\bf q} \to [\hat{\bf z} \times{\bf q}]$. If the incoming
momentum ${\bf q}$ is zero, rotating it by $\pi / 2$ has no effect.
Thus, for ${\bf q} = 0$ we can replace ${\bf S}_X$ in the second diagram
of Fig.~\ref{Fig_Pi_L}c by $\delta {\bf S}_T$, after which it becomes
identical to the totality of diagrams in Fig.~\ref{Fig_Pi_L}a for
$\Pi^*$; therefore,
\begin{equation}
\Pi_L(0, \epsilon) = \Pi^*(0, \epsilon) 
- \frac{\epsilon \epsilon_c}{\Pi^*(0, \epsilon) - \epsilon}.
\label{Pi_L_0}
\end{equation}
As explained in Sec.~\ref{Model}, $\Pi^* \simeq \Pi$; hence, we will
succeed in reproducing Eq.~(\ref{Pi_L}) as soon as we show that
inequality~(\ref{Intermediate_field}) is indeed the relevant condition
for dropping ${\bf S}^*$ in the diagrammatic series for $\Pi_L$. To do
so we develop one step further the argument of Sec.~\ref{Model}.
One-particle irreducible diagrams containing ${\bf S}^*$ have at least
one L-phonon line with typical momentum transfer $k \sim q_c$, see
Fig.~\ref{Fig_Pi_L}d. If $\epsilon_c = 0$ such diagrams are suppressed
by at least a factor of $\mu / \lambda$, as we found previously in
Sec.~\ref{Model}. However, if $\epsilon_c \neq 0$, these diagrams also
generate terms of the order of
\begin{equation}
\sqrt{\epsilon \epsilon_c}\, \frac{1}{H^0_L(k)}
\sqrt{\epsilon \epsilon_c} \sim 
\frac{\epsilon \epsilon_c}{\lambda q_c^{2}},
\label{Additional_term}
\end{equation}
which act as an additional self-energy correction to ${\bf D}_T^*$ in
Fig.~\ref{Fig_Pi_L}c. They can be approximately accounted for if we
replace $\epsilon$ in the argument of $\Pi^*$ in Eq.~(\ref{Pi_L_0}) by a
renormalized value $\tilde\epsilon$,
\begin{eqnarray}
\displaystyle \Pi_L(0, \epsilon) &=& \Pi^*(0, \tilde\epsilon) 
- \frac{\epsilon \epsilon_c}{\Pi^*(0, \tilde\epsilon) - \epsilon},
\label{Pi_L_0_strong}\\
\displaystyle \tilde\epsilon &=& \epsilon
 + C_1 \frac{\epsilon \epsilon_c}{\lambda q_c^{2}},\quad C_1 \sim 1.
\label{tilde_epsilon}
\end{eqnarray}
The difference between $\epsilon$ and $\tilde\epsilon$ can be safely
ignored as long as the inequality~(\ref{Intermediate_field}) is
satisfied, in which case Eq.~(\ref{Pi_L_0}) is {\it asymptotically
exact\/}. In stronger fields Eq.~(\ref{Pi_L_0_strong}) applies, which
gives only the order of magnitude of $\Pi_L$ because of the
uncertainty in the parameter $C_1$.

The meaning of Eqs.~(\ref{Pi_L_0}) and (\ref{Pi_L_0_strong}) can be
elucidated returning to the the original formulation where both
T- and L-degrees of freedom are present:
\begin{mathletters}
\begin{eqnarray}
        \Pi(0, \epsilon) &=& \Pi^*(0, \epsilon),\quad
        \epsilon_c \ll (\lambda / \mu) \Pi_0,
\label{Pi_0}\\
                         &=& \Pi^*(0, \tilde\epsilon),\quad
        \epsilon_c \gg (\lambda / \mu) \Pi_0.
\label{Pi_0_strong}
\end{eqnarray}
\label{Pi_0_B}
\end{mathletters}
We see that $\Pi(0, \epsilon)$ remains magnetic field independent up to
rather high $B$. As explained above, this occurs because the enhancement
of the L-T mixing due to the Lorentz force is strongly impeded by the
disparity of the two elastic moduli.

Let us now evaluate the consequences of the obtained
equations~(\ref{Pi_0}) and (\ref{Pi_0_strong}). In the intermediate
field range, $\Pi_0 \ll \epsilon_c \ll (\lambda / \mu) \Pi_0$, where
Eq.~(\ref{Pi_0}) applies, the consequences are two-fold: the suppression
of $\epsilon_p$, which is well known, and the suppression of the pinning
mode linewidth, which is nontrivial. Indeed, since $\epsilon_p$ becomes
much less than $\Pi_0$ [Eq.~(\ref{epsilon_p_estimate})], it slips into
the low-energy tail of the zero-field phonon spectrum where ${\rm Im}
\Pi \ll \Pi_0$; hence, $\Delta\omega_p \ll \omega_p$. The connection
between the linewidth of the absorption line in {\it strong\/} magnetic
fields and the low-frequency phonon modes in {\it zero\/} magnetic
field, which we just established, is the keystone of the present paper.

The properties of the zero-field soft modes will be discussed in more
detail in Sec.~\ref{Soft_modes} where we argue that they give rise to
the power-law dependence of ${\rm Im}\, \Pi$ on $\epsilon$,
\begin{equation}
  {\rm Im}\, \Pi(\epsilon) \simeq -C_2 \Pi_0
    (\epsilon /\,\Pi_0)^s,\quad \epsilon \ll \Pi_0,
\label{Im_Pi_tail}
\end{equation}
with exponent $s = 3 / 2$ and numerical prefactor $C_2 \sim 1$.
Combining Eqs.~(\ref{Delta_omega_p_def}--\ref{omega_p_def}) and
(\ref{Pi_0_B}--\ref{Im_Pi_tail}), we find for the intermediate-$B$
regime:
\begin{eqnarray}
\displaystyle \sigma_{xx}(\omega) &=& -i
\frac{e^2 n_e \omega}{m \omega_{p 0}^2}
\frac{1 - i f_1(\omega)}{[1 - i f_1(\omega)]^2
 - ({\omega \omega_c}/{\omega_{p 0}^2})^2},
\\
\displaystyle f_1(\omega) &\simeq& \left\{
\begin{array}{cl}
C_2 (\omega / \omega_{p 0})^{2 s},&
\omega \ll \omega_{p 0},
\\
{\rm const},& \omega_{p 0} \ll \omega \ll \omega_c,
\end{array}
\right.
\label{sigma_xx_intermediate}\\
&& \omega_{p 0} \ll \omega_c \ll \omega_{p 0} \sqrt{\lambda / \mu}.
\end{eqnarray}
In this regime the absorption line narrows down according to
Eq.~(\ref{linewidth}), which we reproduce here for convenience:
\begin{equation}
\Delta\omega_p / \omega_p \sim
(\omega_p / \omega_c)^s.
\label{Relative_linewidth_intermediate}
\end{equation}
The high-field regime is described by Eq.~(\ref{Pi_0_strong}). In this
case the relative linewidth is field-independent and is of the order of
\begin{equation}
\Delta\omega_p / \omega_p \sim (\mu / \lambda)^s
\label{Relative_linewidth_strong}
\end{equation}
as given by Eq.~(\ref{linewidth_strong}) and illustrated in
Fig.~\ref{Fig_linewidth}. Although Eq.~(\ref{Pi_0_strong}) was derived
with much less rigor than Eq.~(\ref{Pi_0}), the saturation of
$\Delta\omega_p / \omega_p$ in strong fields is certainly to be
expected. Indeed, in strong fields the dynamics is dominated by the
Lorentz force; therefore, $\epsilon$ must enter in the combination
$\epsilon \epsilon_c$, not by itself. But then a phonon eigenstate,
which for a given (large) $\epsilon_c = \epsilon_c^* = \rho
(\omega_c^*)^2$ has an energy $\epsilon_i$, is also an eigenstate of the
system for larger $\epsilon_c$, with the eigenvalue $\epsilon_i
(\epsilon_c^* / \epsilon_c)$. Hence, as the magnetic field increases,
all relevant eigenfrequencies scale inversely proportional to
$\omega_c$, while the conductivity varies in the self-similar way,
\begin{equation}
\sigma_{xx}(\omega) = \frac{\omega_c^*}{\omega_c}\,
\sigma_{xx}^*\left(\frac{\omega \omega_c}{\omega_c^*}\right),
\label{conductivity_scaling_strong}
\end{equation}
and posseses constant $\Delta\omega_p / \omega_p$. The explicit expression for
$\sigma_{xx}(\omega)$ is similar to Eq.~(\ref{sigma_xx_intermediate}),
\begin{eqnarray}
\displaystyle \sigma_{xx}(\omega) &=& -i
\frac{e^2 n_e \omega}{m \omega_{p 0}^2}
\frac{1 - i f_2(\omega)}{[1 - i f_2(\omega)]^2
 - ({\omega \omega_c}/{\omega_{p 0}^2})^2},
\\
\displaystyle f_2(\omega) &\simeq& \left\{
\begin{array}{cl}
(\omega / \Omega)^{2 s},& \omega \ll \Omega,
\\
{\rm const},& \Omega \ll \omega \ll \omega_c,
\end{array}
\right.
\label{sigma_xx_strong}\\
\Omega &=& C_3 \frac{\omega_{p 0}^2}{\omega_c} \sqrt{{\lambda}/{\mu}},\quad
\omega_c \gg \omega_{p 0} \sqrt{\lambda / \mu},
\end{eqnarray}
where $C_3 \sim 1$.

Formulas~(\ref{sigma_xx_intermediate}) and (\ref{sigma_xx_strong}) are
our final results. They describe the entire lineshape of the pinning
mode both in intermediate and strong magnetic fields. However, their
derivation was presented in diagrammatic rather than physical
terms. Next, we will give an alternative derivation, which elucidates
the physics of the line narrowing and also helps to clarify the
structure of the phonon eigenstates that compose the pinning mode.

We will start with the qualitative picture of localized T-phonons
developed in Sec.~\ref{Model} and add a new ingredient, the stiff
L-degree of freedom. It is clear that an admixture of the L-component
produced by a joint action of the disorder and the Lorentz force makes
phonon eigenstates much more extended in real space. Unlike the softer
T-phonons, the stiffer L-ones cannot be confined in small areas of
size $R_c$. In order to understand the large scale structure of the
eigenstates, we can coarse-grain the system by integrating out the
degrees of freedom on the spacial scales between $a$ and $R_c$. An
insight on the form of the effective Hamiltonian ${\bf H}_L^{eff}$
after the coarse-graining is given by the spectral decomposition of the
matrix element
\begin{equation}
\left \langle {\bf r} \left|
\frac{1}{{\bf H}_T^0 + {\bf S}_T - \epsilon {\bf I}} \right|
{\bf r}^\prime \right\rangle =
\sum\limits_i
\frac{u_{T i}({\bf r}) u_{T i}^*({\bf r}^\prime)}{\epsilon_i - \epsilon}.
\label{Resolvent_decomposition}
\end{equation}
Here $\epsilon_i$ and $u_{T i}({\bf r})$ are the eigenvalues and the
eigenfunctions of localized T-phonons. Since each of $u_{T i}$ is
localized within an area of size $R_c$, we expect that after the
coarse-graining the numerators $u_{T i}({\bf r}) u_{T i}^*({\bf
r}^\prime)$ transform into local operators $R_c^{-2} \delta({\bf r} -
{\bf r}^\prime) \delta({\bf r} - {\bf r}_i)$, where ${\bf r}_i$ is the
center-of-gravity of the $i$th mode. As for the denominators, the
discussion in Sec.~\ref{Model} shows that in weak and intermediate
magnetic fields [defined by the inequality~(\ref{Intermediate_field})],
$\epsilon_i$'s should remain unaffected. Finally, naive averaging of ${\bf
S}_L$ in Eq.~(\ref{H_L}) yields $S_0 {\bf I}$ but the interaction with
high-$k$ T-modes, which can be calculated within the SCBA-type
perturbation theory, renormalizes $S_0$ to $\Pi_0$. The resultant
coarse-grained Hamiltonian is
\begin{equation}
{\bf H}_L^{eff} = {\bf H}_L^0 + \Pi_0 {\bf I} - 
\sum\limits_i
\frac{c_i \Pi_0^2 + \epsilon \epsilon_c}{\epsilon_i - \epsilon}
R_c^2\delta({\bf r} - {\bf r}_i),
\label{H_L_coarse}
\end{equation}
where $c_i \sim 1$ are random and $\epsilon_i \in (0, 2\Pi_0)$. Without
losing essential physics, we can assume that ${\bf r}_i$'s form a
regular square lattice with lattice constant $R_c$.

What are the properties of the obtained lattice model? One of them is
the Lorentzian tails of the distribution function $P(U)$ of the on-site
disorder terms $U_i \equiv (c_i \Pi_0^2 + \epsilon \epsilon_c) /
(\epsilon - \epsilon_i)$. For example, if $\epsilon \lesssim
\epsilon_p$, then
\begin{equation}
P(U) \sim \frac{\Pi_0^2 R_c^2 \nu(\epsilon)}{U^2}, \quad |U| \to \infty,
\label{P_U_tail}
\end{equation}
where $\nu(\epsilon)$ is the T-phonon density of states introduced in
Sec.~\ref{Model}. Clearly, the variance $\langle U^2 \rangle$ of the
on-site disorder is unbounded. One can therefore expect much stronger
disorder effects than in the case of a Gaussian distribution with the
same {\it typical\/} values of $U_i$, i.e., same $\langle \ln |U|
\rangle$. Let $L_L$ be the mean-free path of L-component of the
low-energy phonons. As explained in Sec.~\ref{Model}, Ioffe-Regel
criterion~\cite{Ioffe_60} suggests that $L_L$ is simultaneously their
localization length. For weak Gaussian disorder and unscreened Coulomb
interactions $L_L$ is exponentially large, see
Appendix~\ref{Supersymmetry}. Now we will show that in the presence of
the long tails~(\ref{P_U_tail}), $L_L$ is only as a power-law function
of the disorder strength. The crucial point is that the scattering is
dominated by a few largest $U_i$'s. As soon as we realize this, we can
replace $P(U)$ by the Cauchy distribution
\begin{equation}
P(U) = \frac{1}{\pi} \frac{\Gamma}{(U - U_0)^2 + \Gamma^2},\quad
\Gamma(\epsilon) \sim \Pi_0^2 R_c^2 \nu(\epsilon)
\label{Cauchy}
\end{equation}
on the grounds that it has the same tails. At this point we make a
reasonable assumption that correlations among $U_i$'s at different sites
can be neglected, and arrive at the famous Lloyd model.
The self-energy can now be calculated exactly,~\cite{Lloyd_69}
\begin{equation}
   \Pi_L(\epsilon) = \Pi_0 - U_0(\epsilon) - i \Gamma(\epsilon),
\label{Pi_L_Lloyd}
\end{equation}
while the mean-free path is the solution of the equation $H_L^0(L_L^{-1})
= |{\rm Im}\, \Pi_L|$, which gives $L_L \sim Y / \Gamma$.

To compare with our earlier results we need to know $\nu$ and $U_0$.
The integral in Eq.~(\ref{nu_def}) is determined by $|{\bf k}|
\lesssim q_c$; therefore,
\begin{equation}
\nu(\epsilon) \sim q_c^2 \frac{{\rm Im}\, \Pi^*(\epsilon)}{\Pi_0^2}
\sim \mu^{-1} \left(\frac{\epsilon}{\Pi_0}\right)^s,\quad
\epsilon \lesssim \Pi_0.
\label{nu}
\end{equation}
The estimate of $U_0$ is $U_0 \sim \epsilon \epsilon_c / \Pi_0$.
Combining these together, we find that Eq.~(\ref{Pi_L_Lloyd}) is in
agreement with Eq.~(\ref{Pi_L_0}), which validates our mapping onto
the Lloyd model.

To describe the dominant scattering mechanism in the Lloyd model notice
that ${\rm Im}\, \Pi_L$ comes from the inhomogeneous broadening of the
phonon energies, i.e., the energy shifts $\delta \epsilon$ caused by
on-site disorder $U_i$. For each localized L-phonon the largest energy
shift comes from a site with largest $U_i$ within the localization area
$L_L \times L_L$, which is typically $U_i \sim (L_L / R_c)^2 \Gamma$ for
the Cauchy distribution~(\ref{Cauchy}). Since the amplitude of the
L-phonon wavefunction at the position of the oscillator is $u_L \sim 1 /
L_L$, the resultant energy shift is $\delta \epsilon \sim U_i u_L^2
R_c^2 \sim \Gamma$. The energy shifts from other sites are subdominant;
thus, ${\rm Im}\, \Pi_L \sim \Gamma$, in agreement with the exact result,
Eq.~(\ref{Pi_L_Lloyd}). In the original formulation large $U_i$ come
from the localized T-oscillators whose energy $\epsilon_i$ is almost in
resonance with $\epsilon$, i.e., $ |\epsilon_i - \epsilon| \sim 1 /
\nu(\epsilon) L_L^2$. A few of such resonant scatterers will localize
the L-component on the scale of the mean-free path $L_L$. The proposed
picture resembles the situation in a crystal pinned by strong dilute
impurities, the role of impurities played by the resonant sites.

\begin{figure}
\centerline{
\psfig{file=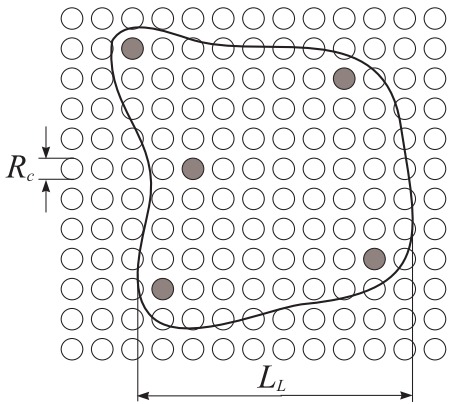,width=2.3in,%
bbllx=108pt,bblly=461pt,bburx=318pt,bbury=651pt}}
\vspace{0.1in}
\setlength{\columnwidth}{3.2in}
\centerline{\caption{Schematic structure of
the localized phonon eigenstate in the crystal with both T- and
L-degrees of freedom present. Circles represent the T-modes
of Fig.~\ref{Fig_localized_T}, the resonant modes are shaded.
\label{Fig_localized_full}
}}
\end{figure}

The structure of a typical phonon is illustrated in
Fig.~\ref{Fig_localized_full}. It involves of the order of $(L_L /
R_c)^2$ T-modes of Fig.~\ref{Fig_localized_T} oscillating with the same
frequency imposed by the mutual coupling mediated by the L-phonon. The
oscillation amplitude is roughly the same for all the T-modes except for
a few resonant ones, where it is much larger. These resonant sites ``drain''
the energy of the L-degree of freedom thereby localizing it.

%
%

Let us now describe the evolution of the phonon eigenstates as the
magnetic field increases. Since the pinning mode frequency goes down,
lower and lower frequency resonant sites are required to effectively
scatter and localize the mixed L-T phonon vibrations. Such sites are
therefore the soft modes, the oscillators of unusually low frequency,
which shape up the tails of the zero-field phonon spectrum (Lifshitz
tails). The soft modes appear due to very rare, peculiar disorder
configurations. The lower the frequency, the mode dilute in real space
they are. Correspondingly, the mean-free path (localization length) of
the magnetophonons becomes larger and larger. The narrowing of the
absorption line is then similar to the motional narrowing phenomenon.
Eventually, in very strong magnetic fields, the enhancement of the L-T
mixing by the Lorentz force start to diminish the frequencies of the
localized T-phonons. At this point slipping of the pinning frequency
$\omega_p$ deeper into the soft-mode tail stops. The wavefunctions of
the phonons cease to change, only their eigenfrequencies continue
decreasing in inverse proportion to the magnetic field. As a result,
$\Delta\omega_p / \omega_p$ remains constant.
 
The physical structure of the pinned phonon modes being clarified, the
only important questions that remain to be considered are the nature of
the soft modes and the derivation of Eq.~(\ref{Im_Pi_tail}). This will
be the subject of the next section.

\section{Soft modes}
\label{Soft_modes}

It is clear that the soft phonon modes must come from rare places where
pinning is unusually weak. The actual calculation of the density of
states $\nu(\epsilon)$ in the low-$\epsilon$ tail is however nontrivial.
The method most suitable for the task seems to be the method of optimal
fluctuation. It has been successfully employed for calculation of the
Lifshitz tails in other disordered systems,~\cite{Lifshitz_book} where
it proves to be asymptotically exact. The method is based on the
following idea. The energy $\epsilon$ of a given eigenstate is
determined in a complicated way by the distribution of the random
potential within the entire area supporting the eigenstate and so there
are many different random potential configurations, which give the same
eigenenergy. If $\epsilon$ is small, any such configuration is
untypical, i.e., a fluctuation of some sort. The method is based on the
assumption that certain fluctuations have much higher
probability than the rest and dominate the quantity of interest,
$\nu(\epsilon)$ in this case. The objective is then to design such an
optimal fluctuation and evaluate its probability.

Let us mention one type of fluctuation, which is not optimal. This is a
configuration where the random potential is strongly suppressed in a
large area $L \times L$, where $L \sim \sqrt{\mu / \epsilon}$. (It may
be useful to recall at this point that we are investigating the soft
modes of a system where L-phonons are integrated out). The probability
of such a fluctuation can be estimated by mentally dividing the area
into uncorrelated blocks of size $R_c \times R_c$ and multiplying
together the probabilities $\sim \epsilon / \Pi_0$ that the random
potential is suppressed within each block. The total probability is
exponentially small $\sim \exp[-(\Pi_0 / \epsilon) \ln (\Pi_0 /
\epsilon)]$. This is a general situation for fluctuations spread over a
large area in real space. Thus, the optimal fluctuation must be of the
smallest possible size. It is easy to see that this size is $R_c$.
Indeed, to get a small phonon eigenenergy $\epsilon$ in a small volume,
the positive ``kinetic energy'' $\mu L^{-2}$ must be accurately
compensated by the negative ``potential energy'' $\langle \delta {\bf
S}_T \rangle_L$. The fluctuations of the latter averaged over the area
$L \times L$ are of the order of $\Pi_0 (R_c / L)$. Thus, for $L \sim
R_c$ the cancellation of $\mu L^{-2}$ can be done by a {\it typical\/}
fluctuation of $\delta{\bf S}_T$ and the probability of such an event
would not have any exponential suppression. Instead, we expect the
power-law dependence of $\nu(\epsilon)$ on $\epsilon$, Eq.~(\ref{nu}).
  
In order to possess a soft mode, the system must have a soft direction,
i.e., a certain collective coordinate $X$ such that the total energy of
the system $E$ as a function of $X$ has a very shallow minimum. Let $X =
0$ be the ground state. In its vicinity $E$ should be Taylor-expandable,
\begin{equation}
                E(X) = \frac{\alpha}{2} X^2 + \beta X^3 + \gamma X^4 + \ldots
\label{E}
\end{equation}
The coefficient $\alpha$ has the meaning of an effective spring constant of
the local oscillator, while the phonon energy $\epsilon$ is determined
by the ratio of the spring constant and the effective mass. The latter
is proportional to the area involved in the oscillations. As we argued
above, this area is of the order of $R_c^2$ for all $\epsilon < \epsilon_p$.
Therefore, $\alpha$ should scale linearly with $\epsilon$ in the
limit $\epsilon \to 0$,
\begin{equation}
                               \alpha = C_4 \epsilon.
\label{alpha_epsilon}
\end{equation}
If we find a way to estimate the probability of such untypically small
$\alpha$, we will succeed in calculating the exponent $s$ of the power-law
function $\nu(\epsilon)$.

This kind of calculation is hardly possible without specifying $X$. Actually,
ascribing the precise meaning to $X$ amounts more or less to designing
the optimal fluctuation. At the moment we do not have a good
understanding how to do it in the 2D case. However, in one dimension the
general structure of the calculation is clear. We are guided by the
earlier work on the subject by Feigelman~{\it et~al.\/}~\cite{Feigelman}
and by Aleiner and Ruzin.~\cite{Aleiner_94}

\begin{figure}
\centerline{
\psfig{file=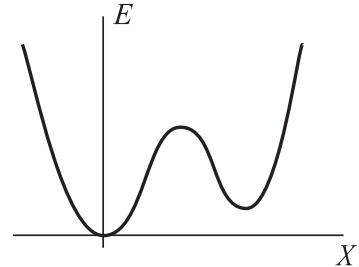,width=1.8in,%
bbllx=215pt,bblly=369pt,bburx=418pt,bbury=524pt}}
\vspace{0.1in}
\setlength{\columnwidth}{3.2in}
\centerline{\caption{Energy of the system as a function of
the collective coordinate $X$.
\label{Energy_landscape}
}}
\end{figure}

In the case of a weakly pinned 1D elastic chain, the role of the desired
collective coordinate $X$ can be played by the elastic displacement
field $u$ at an arbitrary point in the middle of the chain. This point
divides the system into two parts, which communicate only through the
single variable $u$. For any fixed $u$ serving as a boundary condition,
we can find separately the ground state energies $E_<(u)$ and $E_>(u)$
of the left and the right halves of the chain. The ground state of the
whole system is determined by minimizing the sum $E(u) = E_<(u) +
E_>(u)$ with respect to $u$. Both $E_<(u)$ and $E_>(u)$ are periodic
function of $u$ with the period equal to the lattice constant $a$. On
their period they have several (typically, of the order of $a / \xi$)
minima and maxima. The beautiful idea of Aleiner and
Ruzin~\cite{Aleiner_94} (totally overlooked in
Ref.~\onlinecite{Feigelman}) was that the most efficient way to generate
a soft mode is via a frustration, when a maximum of $E_<(u)$ occurs near
a minimun of $E_>(u)$. In this case the second derivative of $E$, i.e.,
$\alpha = E^{\prime\prime}$, can be very small even though each of the
second derivatives of $E_<(u)$ and $E_>(u)$ are typical. The next step
is to show that the probability density distribution $P(\alpha)$ of
$\alpha$ at local minima of $E$ vanishes linearly with $\alpha$ in the
limit $\alpha \to 0$. Indeed,
\begin{eqnarray}
  P(\alpha) &=& \left.\langle \delta(E^{\prime\prime} - \alpha)
                \rangle\right|_{E^\prime = 0}
\nonumber\\
\mbox{} &=& \left \langle \frac{1}{a}
\int\limits_0^a d u \delta(E^{\prime\prime} - \alpha)
         \delta(E^\prime)
\left|\frac{\partial E^\prime}{\partial u}\right| \right\rangle
\nonumber\\
\mbox{} &\to& |\alpha| \langle \delta(E^{\prime\prime}) \delta(E^\prime) \rangle
\nonumber\\
\mbox{} &=& |\alpha| \langle \delta(E_<^{\prime\prime} + E_>^{\prime\prime})
\delta(E_<^\prime + E_>^\prime) \rangle.
\label{P_a}
\end{eqnarray}
The last equation explicitly demonstrates the aforementioned
cancellations between the derivatives of $E_<$ and $E_>$. The value of
the delta-function product average is determined by properties of
typical configurations. Hence, the low probability of having unusually
shallow local minima is entirely due to the first factor on the
left-hand side of Eq.~(\ref{P_a}) and so $P(\alpha)
\propto |\alpha|$. In particular, $P \propto \epsilon$ in the
case of interest~(\ref{alpha_epsilon}). But this is not yet the end of
the story. Following Aleiner and Ruzin~\cite{Aleiner_94} we argue that
the probability of having a shallow {\it global\/} minimum is
additionally suppressed. Indeed, if the coefficient $\beta$ in front of
the the cubic term in Eq.~(\ref{E}) is too large, $E(X)$ would have a
second minimum which is deeper than that at the ground state $X = 0$. To
avoid that $|\beta|$ must not exceed $\gamma \sqrt{\alpha} \propto
\sqrt{\epsilon}$. The typical dependence of the resultant $E(X)$ is
illustrated in Fig.~\ref{Energy_landscape}.

%
%

The total probability density of the optimal fluctuation is therefore
proportional to $\epsilon \cdot \sqrt{\epsilon} = \epsilon^s$ with $s =
3 / 2$, as we claimed. 

It is very likely that $s = 3 / 2$ is in fact the general result
independent of the number of dimensions because the optimal fluctuation
would always have only one soft direction, i.e., the system is
essentially one-dimensional.~\cite{Aleiner_94}

Finally, let us compare our results with those in the literature. One
group of works, by Feigelman {\it et~al.\/}~\cite{Feigelman} has been
already mentioned before. We borrowed from them the idea of splitting
the system into two statistically independent parts. The major mistake
of these authors is overlooking the possibility of frustrations in the
system. In other words, they did not realize that it is not necessary to
require that both $E_<$ and $E_>$ have a small second derivative with
respect to $u$ when only their sum $E_< + E_>$ needs to be so. Another
point of disagreement between us and them is the size of the localized
phonons. Feigelman {\it et~al.\/} take for granted that it is of the
order of $\sqrt{\mu / \epsilon}$, while we gave an argument that it
should be $R_c$, i.e., much smaller.

In another large group of papers,
Refs.~\onlinecite{Fukuyama_77,Fukuyama_78,Chitra_98,Maurey_95},
$\nu_\omega(\omega) \propto \omega^2$ for the density of states, and a
similar dependence, ${\rm Re}\, \sigma_{xx}(\omega) \propto \omega^2$, for
the conductivity were calculated. To facilitate the comparison we should
point out that the conventionally defined phonon density of
states~\cite{Fukuyama_77} $\nu_\omega(\omega)$ is related to our
$\nu(\epsilon)$ by
\begin{equation}
      \nu_\omega(\omega) =  2 \nu(\rho \omega^2) \rho \omega.
\label{nu_omega_nu}
\end{equation}
Hence, our result is ${\rm Re}\, \sigma_{xx}(\omega) \propto
\nu_\omega(\omega) \propto \omega^4$, same as in
Refs.~\onlinecite{Aleiner_94} (see also
Ref.~\onlinecite{Karpov_83}). All the papers in the second group are
explicitly or implicitly based on the SCBA (see Appendix~\ref{SCBA}
for more details). The second power of $\omega$ can be traced to the
unphysical square-root singularity $\nu(\epsilon) \propto |\epsilon -
\epsilon_{\rm th}|^{1/2}$ near a band edge $\epsilon_{\rm th}$
($\epsilon_{\rm th} = 0$ in our case), which is a well-known basic
flaw of the SCBA.~\cite{Lifshitz_book} Note however that even within
the SCBA $\nu(\epsilon)$ conforms to the generic form~(\ref{nu}),
except it corresponds to $s = 1 / 2$ instead of what we believe is the
correct result, $s = 3 / 2$.

\section{Quantum and thermal effects}
\label{Quantum_Thermal}

So far we have neglected any effects of quantum nature or due to a
finite temperature. Some of them will be addressed in this section but
their systematic, in-depth treatment is deferred for future work.

Let us begin with noting that the electrons of the pinned WC constantly
fluctuate around their equilibrium positions. The {\it typical\/} size
of such fluctuations is easy to find. For example, in strong magnetic
fields and at low temperatures ($k_B T \ll \hbar \omega_c$) where all
the electrons are confined to the lowest Landau level, we have
\begin{eqnarray}
\displaystyle \langle {\bf u}^2 \rangle &=& l_B^2
 + \frac{\hbar}{\rho \omega_c} \int\limits_{\bf q}
                       \coth \frac{\hbar\Omega({\bf q})}{2 k_B T}
\nonumber\\
\displaystyle &\mbox{}& \quad\quad\quad \times
\frac{1}{\rho \omega_c \Omega({\bf q})}
\left[\Pi_0 + \frac{\mu({\bf q}) + \lambda({\bf q})}{2} q^2 \right],
\label{u_variance}\\
\displaystyle \Omega({\bf q}) &=& (\rho \omega_c)^{-1}
   \sqrt{[\Pi_0 + \mu({\bf q}) q^2)] [\Pi_0 + \lambda({\bf q}) q^2]}.
\label{omega_mph}
\end{eqnarray}
This formula can be used away from the thermal or quantum melting
transitions, where the phenomenological criterion $\langle {\bf u}^2
\rangle^{1/2} < \varepsilon a$ is satisfied. Here $\varepsilon \sim 0.2$
is the Lindenmann parameter.~\cite{WC_books} Incidentally, at $T =
\Pi_0 = 0$, Eq.~(\ref{u_variance}) gives the variance of the electron
fluctuations in the correlated WC of Lam and Girvin.~\cite{Lam_84}

Let us focus on the range of temperatures $k_B T \ll \hbar\omega_B$
where $\omega_B \sim 4 \pi \mu / m_e \omega_c$ is the magnetophonon
bandwidth (usually, $\hbar\omega_B / k_B \sim 2\,{\rm K}$). At such
temperatures $\langle {\bf u}^2 \rangle^{1/2}$ is totally dominated by
quantum fluctuations and is of the order of $l_B$. We expect that as
long as $\langle {\bf u}^2 \rangle^{1/2}$ is smaller than the
correlation length of the pinning potential $\xi$, which is the smallest
length scale in the classical theory, all the results obtained in the
previous sections acquire at most minor corrections.

Recently Fertig~\cite{Fertig_99} and Chitra~{\it
et~al.\/},~\cite{Chitra_98} studied the opposite limit, $\xi \gg l_B$,
and found a novel dependence of the pinning frequency $\omega_p$ on the
magnetic field, although they disagree with each other on the functional
form of such a dependence. To sort things out we offer a different
perspective on this question. Recall that the classical theory predicts
the $1 / B$-behavior,~\cite{Fukuyama_78}
\begin{equation}
\omega_p \sim \frac{C(0) c}{e \mu \xi^4 B},
\label{omega_p_long} 
\end{equation}
which follows from Eqs.~(\ref{R_c}), (\ref{omega_p_def}), and
(\ref{Pi_0_estimate}). We propose that a reasonably accurate estimate
for $\omega_p$ in the regime $\xi \ll l_B$ can be obtained within a
quasiclassical approximation. The electrons are visualized as compact
wavepackets whose centers of gravity perform classical motion, while the
quantum effects are presumed to be contained in the form-factor of the
wavepakets. This type of quantum effects amount to replacing the bare
pinning potential $U({\bf r})$ by its convolution with the form-factor
$F_e({\bf r})$ of the wavepackets. For small fluctuations $l_B \ll a$
the appropriate form-factor is Gaussian, $F_e({\bf r}) = (\pi \langle
{\bf u}^2 \rangle)^{-1} \exp(- r^2 / \langle {\bf u}^2 \rangle)$.
Upon the convolution, we obtain an effective random potential, with the
correlation length $l_B$ and variance $C(0) \xi^2 / l_B^2$. In view of
such modifications, Eq.~(\ref{omega_p_long}) transforms into
\begin{equation}
\omega_p \sim \frac{e^2}{\mu \hbar^3 c^2} C(0) \xi^2 B^2,
\quad \xi < l_B.
\label{omega_p_short} 
\end{equation}
The quasiclassical approximation is certainly not exact; nonetheless,
when used away from the quantum Hall fractions, at sufficiently low $T$,
and for {\it robust quantities\/} such as $\omega_p$, it should be
qualitatively correct.

The linear in $B$-dependence of $\omega_p$ derived by Fertig [see
Eq.~(44) in Ref.~\onlinecite{Fertig_99}] is at odds with the
$B^2$-dependence found above [Eq.~(\ref{omega_p_short})] and in
Ref.~\onlinecite{Chitra_98}. Although Fertig consider a Poissonian
disorder (potential wells or ``pits'' of size $s < l_B$, energy $\Delta
V$, and the areal density $n_i < 1 / \pi l_B^2$) rather than the
Gaussian one, the real source of the discrepancy is his ascertion that
the elastic distortion at the Larkin length scale is of the order of
$n_i^{-1/2}$. Instead, $l_B$ should be used. Upon this correction, the
formula for $\omega_p$ acquires the above form~(\ref{omega_p_short})
with $C(0) \xi^2$ replaced by $n_i s^4 \Delta V^2$.

\begin{figure}
\centerline{
\psfig{file=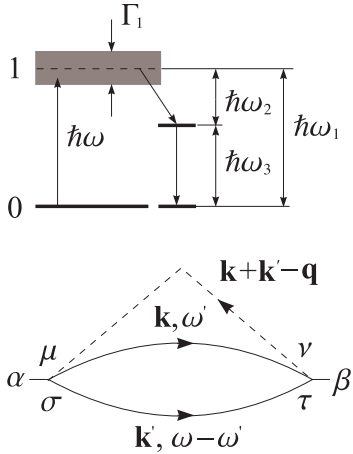,width=1.8in,%
bbllx=215pt,bblly=361pt,bburx=413pt,bbury=619pt}}
\vspace{0.1in}
\setlength{\columnwidth}{3.2in}
\centerline{\caption{Top: low-temperature absorption in
a system with energy levels broadened by anharmonicities.
Bottom: the diagram used for estimating $\Gamma_1$.
\label{Quantum_broadening}
}}
\end{figure}

The finite amplitude of the quantum fluctuations has another important
repercussion: nonvanishing high-order in ${\bf u}$ terms in the pinning
energy. As a result, the absorption line acquires extra broadening
even at $T = 0$. It originates from finite widths $\Gamma_n$ of the
energy levels $n = 1, 2, \ldots$ of the localized magnetophonon
oscillators (the ground state $n = 0$ remains unbroadened). At zero $T$
only the transitions between $n = 0$ and $n = 1$ levels contribute to
the absorption. The level width $\Gamma_1$ of the $n = 1$ state is
determined by the decay of the given magnetophonon into two other
magnetophonon excitations of lower frequencies, $\hbar\omega_1 \to
\hbar\omega_2 + \hbar\omega_3$ (see Fig.~\ref{Quantum_broadening}).

Such a process is always possible in a truly random system, the
necessary three-phonon matrix elements,
%
\[
M_{\alpha \beta \gamma}({\bf r}) = -n_e
\sum_{{\bf K}} K_\alpha K_\beta K_\gamma U({\bf r})
\sin {\bf K} [{\bf r}- {\bf u}^{(0)}({\bf r})],
\]
%
being provided by the cubic anharmonicities. The corresponding
contribution ${\rm Im}\,\Pi_3$ to the imaginary part of the self-energy
can be obtained by evaluating the diagram in
Fig.~\ref{Quantum_broadening},
\begin{eqnarray}
\displaystyle {\rm Im}\,\Pi_3({\bf q}, \omega) &=& -\frac{2 \hbar}{L^2}
\int\limits_{\bf k} \int\limits_{{\bf k}^\prime}
\langle \tilde{M}_{\alpha \mu \sigma} \tilde{M}_{\beta \nu \tau} \rangle
\int \frac{d \omega^\prime}{2 \pi}
\nonumber\\
\mbox{} &\times&
[n_B(\omega^\prime) + n_B(\omega - \omega^\prime) + 1]
\nonumber\\
\mbox{} &\times&
{\rm Im}\, D_{\mu \nu}({\bf k}, \omega^\prime)
{\rm Im}\, D_{\sigma \tau}({\bf k}^\prime, \omega - \omega^\prime),
\label{Im_Pi_3}\\
n_B(\omega) &=& \frac{1}{\exp(\hbar \omega / k_B T) - 1}.
\label{n_B} 
\end{eqnarray}
At $k_B T \ll \hbar\omega_p$ only the frequency interval $0 <
\omega^\prime < \omega$ contributes to the integral, which leads us to
the estimate
\begin{equation}
{\rm Im}\,\Pi_3(\omega_p) \sim {\rm Im}\,
\Pi({\omega_p}/{2})
\frac{l_B^2}{\xi^2} \frac{a^2}{R_c^2}
\frac{|{\rm Im}\,\Pi({\omega_p}/{2})|}{\Pi_0}.
\label{Im_Pi_3_estimate} 
\end{equation}
This formula is derived assuming that $\xi \gg l_B$; otherwise, in the
spirit of the quasiclassical approximation, we have to replace $\xi$ by
$l_B$. Even in that case ${\rm Im}\,\Pi_3(\omega_p)$ is smaller
than the previously found ${\rm Im}\,\Pi(\omega_p)$ by a large factor
$\sim (R_c / a)^2 (\lambda / \mu)^s$. At higher temperatures, the anharmonic
effects are enhanced by a factor of $k_B T / \hbar \omega_p$ originating
from the Bose distribution functions $n_B$ in Eq.~(\ref{Im_Pi_3}). They
should start to affect the linewidth when $k_B T \sim (R_c / a)^{7/2} \hbar
\omega_p$ (cf.~Ref.~\onlinecite{Comment_on_Yi}).

%
%

A word of caution is in order. The diagram in
Fig.~\ref{Quantum_broadening} does not contain vertex corrections, which
contain information about, e.g., the soft modes. The soft modes are
crucial for the dynamical response in strong magnetic fields and at the
same time are potential sources of much larger anharmonicities. Thus,
the response of a weakly pinned quantum WC may prove to be more
nontrivial.

Concluding this section, we would like to re-emphasize that for the
problem studied in the main part of the paper (zero-temperature
linear-response of a classical WC), the anharmonicities are not
important.

\section{Comparison with experiment}
\label{Discussion}
 
In this section we will attempt to analyze the most recent
experimental data~\cite{Li_98,Beya} in the light of our understanding
of the pinning mode. Let us first discuss the position of the pinning
resonance. 
 
From Eqs.~(\ref{omega_p_long}) and (\ref{omega_p_short}) we see that
depending on the type of disorder, $\omega_p$ can either decrease as $1
/ B$ or increase as $B^2$ when the magnetic field
increases.~\cite{Chitra_98} Weak magnetic field dependence of $\omega_p$
found in the experiments~\cite{Li_98,Beya} leads us to conclude that
$\xi$ remains {\it of the order of\/} $l_B$ in the limited range
$10$--$15\,{\rm T}$ of the magnetic fields where the pinning mode was
detected, i.e., $\xi \approx l_B \approx 75{\rm \AA}$. The source of
such a short-range random potential is likely to be the roughness of the
heterostructure interface as suggested by Fertig.~\cite{Fertig_99} The
experimental value of $\omega_p \approx 8 \times 10^9\, s^{-1}$ for $n_e
= 5.4 \times 10^{11}\,{\rm cm}^{-2}$ can then be reproduced with
reasonable values of fitting parameters (see Sec.~\ref{Quantum_Thermal})
$s = 30\,{\rm\AA}$, $\Delta V = 4.3\,{\rm K}$, and $n_i = 4 \times
10^{11}\,{\rm cm}^{-2}$. Unfortunately, the disorder parameters are
poorly known, and so we have to use $\omega_p$ to estimate the disorder
characteristics, not the other way around. For example, we can solve
Eq.~(\ref{omega_p_short}) for the rms amplitude $\sqrt{C(0)}$ of the
pinning potential, which gives a value of the order $0.2\,{\rm K}$.

Let us now discuss the density dependence of $\omega_p$. Explicitly, it
enters Eq.~(\ref{omega_p_short}) only through the shear modulus
$\mu(n_e)$. The latter should behave as $\mu \propto n_e^{3 / 2}$ (see
Sec.~\ref{Statics}) away from the quantum Hall fractions, e.g., the
$\frac13$-filling. If $C(0)$ is $n_e$-independent, we therefore expect
$\omega_p \propto n_e^{-3/2}$. Such a dependence was indeed observed by
Li~{\it et~al.\/}~\cite{Li_98} for the concentration of holes ($p$-type
samples were used) in the range $3$--$5 \times 10^{10}\,{\rm cm}^{-2}$,
which corresponds to the filling factor range $0.1$--$0.16$. At lower
concentrations the pinning frequency continued growing as $n_e$
decreased but less rapidly. This comes presumably from the fact that the
pinning is no longer weak at such low $n_e$: one can verify that $R_c$
is approximately $6 a$ at the highest densities, but approaches $a$ (the
lattice constant) at the lowest densities used in
Ref.~\onlinecite{Li_98}. (In Ref.~\onlinecite{Beya} the density
dependence of $\omega_p$ was not investigated).

The relationships among the empirical values of $\omega_p$ and
$\Delta\omega_p$ and the experimental parameters ($n_e$ and $B$) provide
another means to test the agreement between the theory and the
experiment. To this end we rewrite Eq.~(\ref{linewidth_strong}) (with $s
= 3/2$) in the following way:
\begin{equation}
Q \equiv \frac{\omega_p}{\Delta\omega_p} \sim \left[
\left(26 \frac{e c}{\kappa}
\frac{n_e^{3/2}}{B f_{pk}}\right)^{1/2} - 5\right]^{3 / 2},
\label{Q} 
\end{equation}
where $f_{pk} = \omega_p / (2 \pi)$ is the pinning frequency in cycles
per second. The largest quality factor $Q \approx 8$ reported by Li~{\it
et~al.\/} was achieved for $n_e = 5.4 \times 10^{10}\,{\rm cm}^{-2}$,
$B = 13\,{\rm T}$, where $f_{pk}$ was measured to be $1.4\,{\rm GHz}$.
In Ref.~\onlinecite{Beya} $Q \approx 45$ was found for comparable $n_e$,
$B$, and $f_{pk}$. Our theoretical estimate from Eq.~(\ref{Q}) is $Q
\approx 260$ and far exceeds both. We should point out, however, that
specific details of the experimental setup become important for
observability of such a high $Q$. Henceforth we focus on the work of
Li~{\it et~al.\/},~\cite{Li_98} since most of their data is available
in the published record.

\begin{figure}
\centerline{
\psfig{file=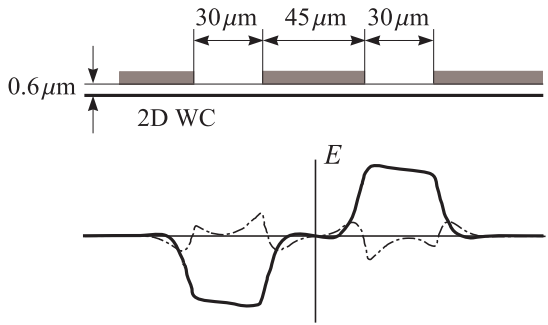,width=2.8in,%
bbllx=128pt,bblly=358pt,bburx=465pt,bbury=559pt}}
\vspace{0.1in}
\setlength{\columnwidth}{3.2in} \centerline{\caption{Top: schematics of
the microwave setup used by Li~{\it et~al.\/} The transmission line is
shaded, the 2D electron (or more precisely, {\it hole\/}) gas is
represented by the thick horizontal line underneath (adapted from
Ref.~\protect\onlinecite{Li_98}). Bottom: typical form of the
electric field distribution in the plane of the
WC.~\protect\cite{Fogler_private} Solid line is the signal in phase with
the source, dashed-dotted line is the signal in quadrature.
\label{Fig_CPW}
}}
\end{figure}

Because of the finite distance ($30\,\mu{\rm m}$) between the plates of
the coplanar microwave waveguide (see Fig.~\ref{Fig_CPW}), these were
not the ideal ${\bf q} = 0$ measurements. We estimate that the spread of
${\bf q}$'s imposes un upper bound of about $30$ for the largest
observable $Q$. The remaining discrepancy seems to be rooted in
finite-temperature and to a lesser degree finite incident power effects
ignored in the derivation of Eq.~(\ref{Q}). In the experiments, the
linewidth was decreasing roughly linearly as the temperature varied from
$T = 200\,{\rm mK}$ to $T = 50\,{\rm mK}$ with other parameters held
fixed. Below $50\,{\rm mK}$ the linewidth appeared to reach saturation,
but only at the nominal level of incident microwave power ($80\,{\rm
pW}$). When lower power levels were used, the line continued sharpening
up. At the lowest experimental temperature of $25\,{\rm mK}$, the
linewidth did not show any signs of saturation (this time as a function
of power) even upon ten-fold input power reduction, which was near the
sensitivity limit.

%
%

Several sources of the thermal broadening can be envisioned: (a)
thermally excited single-particle excitations of the
WC,~\cite{Normand_92} (b) phonons of the host semiconductor, (c) the
anharmonisms of the collective modes of the pinned WC, and perhaps some
others.~\cite{Fertig_99}

The classical activation energy of vacancies and interstitials is of the
order of $0.5\,{\rm K}$ at the densities studied, and mechanism (a)
should be frozen out at $25\,{\rm mK}$. Furthermore, for the WC of low
density the bandwidth of such excitations is very narrow and so they are
easily localized by the random potential.

The weak electron-phonon coupling in GaAs and small phonon phase space
at the frequencies involved ($\omega_p$) are likely to render the
mechanism (b) inefficient.

The preliminary estimate of the broadening due to mechanism (c) was given
in the previous section. It is too small to explain the experimental
observations. We speculate that stronger anharmonic effects and a better
agreement with the experiments may be obtained if one properly accounts
for untypically large anharmonicities at the locations of the soft
modes, which as we showed in this paper, play a very important role in
the response.

It is also quite possible that the actual pinning potential is more
complicated than the one studied here. In reality, it can be
due to a combination of the interface roughness~\cite{Fertig_99} and
dilute residual ions in the vicinity of the WC plane.~\cite{Ruzin_92} In
this case $\omega_p$ may be determined by the former, while $Q$ could be
limited by the latter. Such more complicated models as well as the
dynamic response of a pinned quantum WC are interesting subjects
awaiting further investigation.

It would be also interesting to investigate if the discussed phenomena
appear in the conventional charge-density waves (see
Sec.~\ref{Introduction}). These materials are three-dimensional, and
some important modifications of the present theory may be needed. There
are other complications such as an extra dissipation due to uncondensed
quasiparticles but those can be suppressed by lowering the temperature.
In any case, it would be remarkable if narrow pinning modes could be
produced in these materials simply by applying a sufficiently strong
magnetic field.

Our theory should literally apply to the pinning mode of a WC formed
by electrons on solid hydrogen (see the second book cited under
Ref.~\onlinecite{WC_books}). We hope that this paper will encourage
further experiments on these or other numerous systems where the
pinning manifests itself.

\acknowledgements

This research is supported by US Department of Energy Grant
No.~DE-FG02-90ER40542 and NSF Grant No.~DMR-9802468. We thank R.~Chitra,
Mark Dykman, Lloyd Engel, Herb Fertig, Gabi Kotliar, Chi-Chun Li, Leonid
Pryadko, Misha Raikh, Dan Tsui, and Valerii Vinokur for useful
discussions. M.~M.~F. is grateful to Tito Williams for providng a copy
of Ref.~\onlinecite{Beya}.

\appendix

\section{Model calculation for L-L scattering}
\label{Supersymmetry}

In order to evaluate the importance of the L-L scattering channel,
let us consider the model action
\begin{eqnarray}
A_L(\epsilon) &=& \frac12 \int\limits_{\bf q}
u_L^*({\bf q}) [H^0_L({\bf q}) - \epsilon] u_L({\bf q})
\nonumber\\
\mbox{} &+& \frac12 \int\limits_{{\bf r}}
u_L^*({\bf r}) S_L({\bf r}) u_L({\bf r}).
\label{A_L}
\end{eqnarray}
where $S_L$ is a local operator of a Gaussian white-noise type
with parameters
\[
\langle S_L \rangle = \Pi_0,\quad
\langle S_L({\bf r}_1) S_L({\bf r}_2)\rangle 
= \Pi_0^2 - V^{\prime\prime}(0) \delta({\bf r}_1 - {\bf r}_2).
\]
Our objective is to calculate the disorder-averaged Green's function
\[
D_L({\bf q}, \epsilon) \equiv \frac{i}{L^2}
\left\langle \frac{\int {\cal D} u_L {\cal D}u_L^*
u_L({\bf q}) u^*_L({\bf q}) e^{-i A_L(\epsilon)}}
{\int {\cal D} u_L {\cal D} u_L^* e^{-i A_L(\epsilon)}}
\right\rangle,
\]
%
where an infinitesimal imaginary correction is assumed to be included
via $\epsilon \to \epsilon + i 0$ to make the integrals convergent.

Using the ``supersymmetry'' technique,~\cite{Efetov_book} $D_L$ can be
represented in the form
%
\[
D_L = \frac{i}{L^2}
\left\langle \int {\cal D}{\bf \phi} {\cal D}{\bf \phi}^*
u_L u^*_L e^{-i A_s[\phi]}
\right\rangle.
\]
%
where $\bbox{\phi}_i^\dagger = [u^*_L\: v^*_L]$ is a supervector,
$v_L({\bf q}, \omega)$ is an auxillary fermionic field, and
$A_s[\bbox{\phi}]$ is the sypersymmetric Eucledian action
\begin{eqnarray}
A_s &=& \frac12 \int\limits_{\bf q}
\bbox{\phi}^\dagger({\bf q}) (Y q + \Pi_0 - \epsilon)
\bbox{\phi}({\bf q})
\nonumber\\
\mbox{} &+& \frac{i}{8} V^{\prime\prime}(0) \int\limits_{\bf r}
[\bbox{\phi}^\dagger({\bf r}) \bbox{\phi}({\bf r})]^2.
\label{A_s}
\end{eqnarray}
Note that the coefficient in front of the quartic term is imaginary,
which makes the action non-Hermitian. The sign of the imaginary part is
such that excitations above the vacuum state $\bbox{\phi} \equiv 0$
decay after travelling a certain distance. This distance is simply the
phonon mean free path.

It is easy to establish some of the properties of the propagator $D_L$
right away. Since the field $\bbox{\phi}$ is massive for $\epsilon
\lesssim \Pi_0$ at the tree level, the imaginary part of the
propagator $D_L$ is small at such $\epsilon$. However, ${\rm Im}\, D_L$
should grow rapidly as soon as $\epsilon$ crosses the $\epsilon = \Pi_0$
threshold. Below we are going to verify this by explicit calculations.
Although our methods do not work in the immediate vicinity of $\epsilon
= \Pi_0$, the expressions obtained for $\epsilon < \Pi_0$ and $\epsilon
> \Pi_0$ can be smoothly matched onto each other.

Rescaling the fields in Eq.~(\ref{A_s}) by $\sqrt{2 / Y}$,
we obtain the action with the Lagrangian
\begin{equation}
{\cal L} = \bbox{\phi}^\dagger
\hat{K} \bbox{\phi}
+ \frac{M}{2} \bbox{\phi}^\dagger \bbox{\phi}
- i \frac{g}{4} (\bbox{\phi}^\dagger \bbox{\phi})^2,
\label{Lagrangian}
\end{equation}
where $\hat{K}$ is the operator, which corresponds to multiplication by
$|{\bf q}|$ in the momentum representation, $M = 2 (\Pi_0 - \epsilon) /
Y$ plays the role of mass, and $g \equiv -2 V^{\prime\prime}(0) / Y^2$
is the bare coupling constant. The coupling constant is dimensionless,
and it is possible to show that the field theory is renormalizable. Let
us derive the Wilson-style renormalization group (RG) equations for the
renormalized parameters $g_R$ and $M_R$. As usual, it is done by
successive integrations over narrow momentum shells $\Lambda(l + \delta
l) < q < \Lambda(l)$ where $\Lambda(l) = \Lambda_c e^{-l}$ and $\Lambda_c
\sim 1 / a$ are the running and the bare high-momentum cutoffs,
respectively.

\begin{figure}
\centerline{
\psfig{file=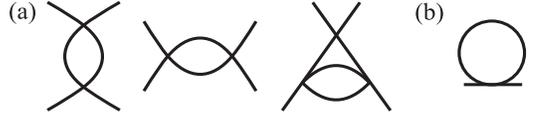,width=2.7in,%
bbllx=103pt,bblly=413pt,bburx=433pt,bbury=485pt}}
\vspace{0.1in}
\setlength{\columnwidth}{3.2in}
\centerline{\caption{
Diagrams for the (a) coupling constant and (b) mass term.
\label{One_loop}
}}
\end{figure}

Consider the effect of such an integration on $g_R$ first. Initially,
$g_R$ remains small and we need to take into account only the three
lowest order in $g_R$ diagrams (see Fig.~\ref{One_loop}a) to find
\begin{eqnarray}
&& \displaystyle g_R(l + \delta l) = g_R(l) + \frac{2}{3} g_R^2
\int\limits_{e^{-\delta l} \Lambda < q < \Lambda}\!\!
\frac{d^2 q}{(2 \pi)^2} \frac{1}{q + M_R}
\nonumber\\
&& \displaystyle\mbox{} \times \left[
  \frac{1}{|{\bf q} + {\bf s}| + M_R}
+ ({\bf s} \to {\bf t}) +  ({\bf s} \to {\bf u})
\right],
\label{Delta_g_R}
\end{eqnarray}
where $S = {\bf s}^2 = ({\bf q}_1 - {\bf q}_2)^2$, $T = {\bf t}^2 =
({\bf q}_1 - {\bf q}_3)^2$, and $U = {\bf u}^2 = ({\bf q}_1 -
{\bf q}_4)^2$ are the Mandelstam variables with ${\bf q}_i$, $i = 1,
\ldots, 4$ being the incoming momenta. Similarly, the tadpole diagram
shown in Fig.~\ref{One_loop}b determines the variation of the
renormalized mass $M_R$,
\begin{equation}
M_R(l + \delta l) \simeq M_R(l) - \frac{g_R}{2}
\int\limits_{e^{-\delta l} \Lambda < q < \Lambda}\!\!
\frac{d^2 q}{(2 \pi)^2} \frac{1}{|{\bf q}| + M_R}.
\label{Delta_M_R}
\end{equation}
%

%
%

In addition to $g_R$ and $M_R$, the kinetic term $\hat{K}$ also gets
renormalized. We will neglect such an effect because it is of higher
(second) order in $g_R$. Consequently, the spectral function ${\rm Im}\,
D_L$ is determined just by the mass,
\begin{equation}
 {\rm Im}\, D_L(q, \epsilon) = \frac{2}{Y}
\frac{{\rm Im}\, M_R}{(q + {\rm Re}\,M_R)^2 + ({\rm Im}\, M_R)^2}.
\label{D_L_M_R}
\end{equation}
In this equation $M_R = M_R(q, \epsilon)$ stands for the renormalized mass
at such large negative $l$ where the RG flow eventually stops either
because $q \sim \Lambda(l)$ or because $|M_R| \sim \Lambda(l)$. 
Below we consider exclusively $q = 0$ case. 

From Eq.~(\ref{Delta_g_R}) we find that in the limit $|{\bf q}_i|, |M_R|
\ll \Lambda$ function $\beta \equiv \partial g_R / \partial l$ depends
only on $g_R$,
\begin{equation}
           \beta = \frac{1}{\pi} g_R^2,
\label{beta_function}
\end{equation}
and so the RG flow equation is easy to solve:
\begin{equation}
           g_R = \frac{\pi}{l_L - l}, \quad l_L \equiv \frac{\pi}{g}.
\label{running_g}
\end{equation}
At $l \approx l_L - \pi$, $g_R$ becomes of the order
of one, and the theory enters the strong coupling regime. The
corresponding spatial scale is $L_{L-L} \sim \Lambda_c^{-1} \exp(\pi /
g)$. Since the strength of coupling maps to the strength of the
effective disorder in the original formulation, $L_{L-L}$ has the
meaning of the L-phonon localization length (due to L-L scattering).
Note that $g_R$ seems to exhibit a divergence (Landau pole) at $l =
l_L$. This is an artifact of the one-loop RG. We expect that instead
$g_R$ flows towards its fixed point value of the order of one.
 
The other scaling equation, for $M_R$, can be deduced from
Eq.~(\ref{Delta_M_R}),
\begin{equation}
\gamma_M \equiv \frac{\partial M_R}{\partial l}
= -\frac{g_R(l)}{4 \pi} \frac{\Lambda^2(l)}{\Lambda(l) + M_R - i \delta}.
\label{gamma_function}
\end{equation}
The solution cannot be not expressed in elementary functions but we can
describe its asymptotics.

\vspace{12pt}
\noindent 1. $(g / 4 \pi) \Lambda_c  = M_* \ll |M| \ll \Lambda_c$.
\vspace{12pt}

Here the renormalization of the coupling constant is not important
because the RG flow is effectively cut off at $l \sim \ln(\Lambda_c /
|M_R|)$ before $g_R$ manages to get large.
Equation~(\ref{gamma_function}) reduces to the SCBA equation
\begin{equation}
 M_R = M - \frac{g}{4 \pi} \int\limits_{\Lambda}^{\Lambda_c} \!\!
\frac{d q q}{q + M_R - i \delta},
\label{M_R_SCBA}
\end{equation}
with the approximate solution  
\begin{mathletters}
\label{Case_1}
\begin{eqnarray}
{\rm Re}\, M_R &\simeq& M - M_*,
\label{Re_M_1}\\
{\rm Im}\, M_R &\simeq& (g / 4) \Theta(-{\rm Re}\, M_R)
{\rm Re}\, M_R,
\label{Im_M_1}
\end{eqnarray}
\end{mathletters}
where $\Theta(x)$ is the Heaviside step-function. A word of caution is
in order here. The imaginary part of $M_R$ vanishes at positive ${\rm Re}
M_R$ only at the level of the one-loop RG. Nonperturbative methods
indicate that ${\rm Im}\, M_R$ is nonzero albeit exponentially small
(see below).

\vspace{12pt}
\noindent 2. $L_{L-L}^{-1} \ll |M - M_*| \ll M_*$.
\vspace{12pt}

Now instead of Eq.~(\ref{Case_1}) we have
\begin{mathletters}
\label{Case_2}
\begin{eqnarray}
&& \displaystyle {\rm Im}\, M_R \simeq \frac14 \frac{\pi}{l_L - l_*}
\Theta(-{\rm Re}\, M_R) {\rm Re}\, M_R,
\label{Im_M_2}\\
&& \displaystyle {\rm Re}\, M_R \simeq \left(\frac{l_L}{l_L -
l_*}\right)^{1 / 4}
(M - M_*),
\label{Re_M_2}\\
&& \displaystyle l_* = \ln\left(-\frac{\Lambda_c}{{\rm Re}\, M_R}\right).
\end{eqnarray}
\end{mathletters}

The above comment about small but nonzero ${\rm Im}\, M_R$ at ${\rm Re}
M_R > 0$ applies here as well. 

\vspace{12pt}
\noindent 3.  $|M - M_*| \lesssim L_{L-L}^{-1}$.
\vspace{12pt}

Since the one-loop RG cannot be trusted beyong the point $l = l_L - \pi$,
we choose to terminate the RG procedure at this stage.
Unfortunately, this leaves us with a theory which (a) does not have
any small parameters and (b) includes not only the original quartic
but also higher order interaction terms generated by the RG. However, on
physical grounds we can expect that in this, essentially massless,
limit, the mass scale must be set by the running cutoff $\Lambda(l_L)$,
i.e.,
\begin{equation}
                {\rm Im} M_R \sim \Lambda_c g^{-1 / 4} e^{-\pi / g}.
\label{Im_M_3}
\end{equation}
This can be compared with the solution of the SCBA equation~(\ref{M_R_SCBA})
in the massless limit (${\rm Re} M_R = 0$),
\begin{equation}
                {\rm Im} M_R^{\rm SCBA} \sim \Lambda_c e^{-4 \pi / g}.
\label{Im_M_SCBA}
\end{equation}
Clearly, the SCBA misses the factor of four in the exponential.

As mentioned above, RG-enhanced perturbation theory formulas~(\ref{Im_M_1})
and (\ref{Im_M_2}) do not work at low energies. Nonvanishing
${\rm Im}\, M_R$ can be detected using the nonperturbative method of
constrained instantons.~\cite{Fogler_private} The details of the calculation
will be presented elsewhere. Here we only quote the result,
\begin{equation}
{\rm Im}\,M_R \simeq \frac{4 \pi e \sqrt{6 \pi}}{9 g_R^2} 
\ln\left(\frac{\Lambda_c g_R}{{\rm Re}\,M_R}\right) \Lambda_c e^{-\pi /
g},
\label{Im_M_R_instanton}
\end{equation}
which is valid for ${\rm Re} M_R \gg 1 / L_{L-L}$, $g_R \ll 1$.
Equations~(\ref{Im_M_3}) and (\ref{Im_M_R_instanton}) match for $|{\rm
Re} M_R| \sim 1 / L_{L-L}$. Therefore, we expect that the spectral
function ${\rm Im} D_L$ exhibits a quasi-Lorentzian peak centered at
$\epsilon_p = Y M_* \simeq \Pi_0$ of exponentially small width
\begin{equation}
\Delta \epsilon_p = \frac{Y}{2}\, {\rm Im} M_R(0, \epsilon_p)
\sim \Lambda_c Y g^{-\omega} e^{-\pi / g}.
\end{equation}
%

\section{Dynamical response within the SCBA}
\label{SCBA}

Within the SCBA the self-energy $\Pi_{\alpha\beta}(\omega, q)$
is diagonal, $\Pi_{\alpha\beta}(\omega, q) = \delta_{\alpha\beta} \Pi$
and weakly $q$-dependent for $q a \ll 1$. $\Pi$ is to be found from
the self-consistency equation~(\ref{Pi_SCBA}) which we reproduce
here for convenience,
\begin{equation}
\Pi(\omega) =  S_0 + V^{\prime\prime}(0) \int\limits_{\bf k}
{\rm tr}\,D({\bf k}, \omega).
\label{Pi_SCBA_Appendix}
\end{equation}
As discussed in Secs.~\ref{Model} and \ref{L_scattering}, the SCBA
applies only at relatively high frequencies. Moreover, in strong
magnetic fields and near the pinning frequency $\omega \sim \omega_p$,
it is not even qualitatively correct. One may ask why we bother
elaborating on this faulty approximation scheme here. The reason is as
follows. We discovered that as far as the dynamical response is
concerned, all alternative theories suggested so
far~\cite{Fukuyama_78,Normand_92,Fertig_99,Chitra_98} are merely
different forms of the SCBA. Thus, we deemed that it would be helpful
to expose their interrelationship and correct a few calculational
errors.

Let us briefly summarize the results obtained by previous authors.
Fukuyama and Lee~\cite{Fukuyama_78} concluded that $\Delta\omega_p \sim
\omega_p$ but did not present the details of the calculation.
Perturbative analysis of Fertig~\cite{Fertig_private} yields
an exponentially small $\Delta\omega_p / \omega_p$ (for weak disorder).
Gaussian variational replica method (GVM) of Chitra~{\it et
al.\/}~\cite{Chitra_98} reduces to the SCBA at nonzero frequencies, and
when analyzed further, predicts a power law dependence of
$\Delta\omega_p$ on the disorder strength. Finally, the SCBA analysis in
the Appendix of Ref.~\onlinecite{Normand_92} leads to yet another
dependence. One reason for such a disarray of conflicting results is the
extreme sensitivity of the solution $\Pi(\omega)$ of
Eq.~(\ref{Pi_SCBA_Appendix}) to the precise value of the dimensionless
parameter $C \equiv -V^{\prime\prime}(0) / (4\pi\mu \Pi_0)$, where
$\Pi_0 \equiv \Pi(0)$. In principle, $C$ is fully determined by $S_0$,
but $S_0$ is known only approximately. A strong partial cancellation of
$S_0$ by the second term in Eq.~(\ref{Pi_SCBA_Appendix}) leaves us only
the order of magnitude estimate $C \sim 1$. One could argue, as Fukuyama
and Lee did in analogous situation,~\cite{Fukuyama_77} that $C = 1$ is
the ``best'' choice because others lead to various physical absurdities.
For example, if $C > 1$, then $\Pi(\omega)$ acquires a finite imaginary
part at complex $\omega$, which means that the ground state is unstable;
if $C < 1$, then $\Pi(\omega)$ has an imaginary part only above some
threshold frequency $\omega_{\rm th} > 0$, i.e., the phonon spectrum has
a gap, which also appears to be unphysical. The truth is, of course,
that the SCBA is simply unable to correctly describe the properties of
the $\omega \to 0$ tail; therefore, it cannot be relied upon for
selecting a ``good'' value of $C$. More sensible approach is to
investigate a certain range of $C$ around unity, hoping that certain
quantities depend on $C$ only weakly. Alas, the linewidth
$\Delta\omega_p$ is not one of such quantities.

We found that the results of Chitra~{\it et al.\/} are recovered for the
``Fukuyama-Lee choice,'' $C = 1$. (Therefore, these two groups of
authors should have obtained the same results). A slight reduction of
$C$ from unity is sufficient to cross over to very different predictions
of Fertig. Let us now demonstrate this in more detail.

Using the definition of $C$ and Eq.~(\ref{Pi_SCBA_Appendix}), we obtain
\begin{equation}
\Pi(\omega) - \Pi_0 = 2 \mu C \Pi_0 \int\limits_0^\infty \!
d k k \left[D_T(k, 0) - D_T(k, \omega) \right].
\label{Pi_SCBA_dif}
\end{equation}
Let us now describe the solution $\Pi(\omega)$ of this
equation for different $C$ in the weak pinning regime $R_c \gg a$, which
corresponds to the inequality $\alpha \equiv \sqrt{\Pi_0\mu} / Y 
= \mu / \lambda \ll 1$.

\vspace{12pt}
\noindent 1. $0 < 1 - C \lesssim \alpha$.
\vspace{12pt}

In this case
\begin{equation}
\Pi(\omega) \simeq C \Pi_0 + i \Pi_0
\sqrt{(\omega / \tilde{\omega}_p)^2 - (1 - C)^2},
\label{Pi_SCBA_1}
\end{equation}
where $\tilde{\omega}_p = \omega_p / (\pi \alpha)^{1 / 2}$ and $\omega_p
= \Pi_0 / \rho \omega_c$. As one can see, the threshold frequency is
$\omega_{\rm th} = (1 - C)\tilde{\omega}_p$. For $C = 1$
Eq.~(\ref{Pi_SCBA_1}) coincides with that of
Ref.~\onlinecite{Chitra_98}. Strictly speaking, Eq.~(\ref{Pi_SCBA_1})
describes the solution of Eq.~(\ref{Pi_SCBA_dif}) only for $\omega \ll
\omega_p$. Near the resonance, $\omega \sim \omega_p$, it is off by a
numerical factor of the order of unity. In principle, this factor can
also be calculated. We limit ourselves to the order of magnitude
estimate ${\rm Im}\,\Pi(\omega_{p}) \sim \alpha^{1/2} \Pi_0$, which leads to
$\Delta\omega_p / \omega_p \sim \alpha^{1 / 2} = (\mu / \lambda)^{1 / 2}$.
Although this result is of the same form as
Eq.~(\ref{Relative_linewidth_strong}), with $s = 1 / 2$, we believe that
$s = 3 / 2$ is correct, see Secs.~\ref{L_scattering} and \ref{Soft_modes}.

\vspace{12pt}
\noindent 2. $1 - C \gg \alpha$.
\vspace{12pt}

Equation (\ref{Pi_SCBA_1}) still applies as long as $\omega_{p 0} -
\omega$ is positive and not too small. At larger frequencies,
$\Pi(\omega)$ receives additional contribution from the pole $k_*$ of
$D_T$ in the complex plane of $k$, which moves close to the real axis.
The resonance linewidth $\Delta\omega_p$ is exponentially small,
\begin{equation}
\Delta\omega_p  =  \frac{{\rm  Im}\, \Pi(\omega_p)}{\rho \omega_c}
 \sim \omega_p e^{-F / 4 \alpha^2},
\label{Delta_omega_p_SCBA_2}
\end{equation}
where $F \simeq \Pi(\omega_{\rm th}) / (C \Pi_0) - 1$. The threshold
frequency $\omega_{\rm th}$ is only slightly below $\omega_p$, by an
amount of the order of $\Delta\omega_p$. To be exact $\omega_{\rm th}$
can be found from the condition that the solution of
Eq.~(\ref{Pi_SCBA_dif}) is also the solution of the ``derivative'' of
this equation with respect to $\Pi$,
\begin{equation}
1 = 2 \mu C \Pi_0 \int\limits_0^\infty
\frac{ d k k [\epsilon \epsilon_c + (\Pi + Y k)^2]}
{[(\Pi + \mu k^2)(\Pi + Y k) - \epsilon \epsilon_c]^2}.
\label{Pi_SCBA_dif_derivative}
\end{equation}
At frequencies somewhat higher than $\omega_p$, ${\rm Im}\, \Pi(\omega)$
is dominated by the aforementioned pole at $k_* = (\epsilon \epsilon_c -
\Pi^2) / \Pi Y$ near the real axis, $|{\rm Re}\, k_*| \ll
|{\rm Im}\, k_*|$. ${\rm Im}\, \Pi$ is given simply by the
residue of this pole,~\cite{Normand_92}
\begin{equation}
{\rm Im}\,\Pi(\omega) \simeq 2 \pi \alpha^2 C \Pi_0
\left(\frac{\omega^2}{\omega_p^2} - 1\right). 
\label{Pi_SCBA_2}
\end{equation}
In conclusion, we would like to reiterate that neither of
Eqs.~(\ref{Pi_SCBA_1}), (\ref{Delta_omega_p_SCBA_2}), or
(\ref{Pi_SCBA_2}) describes the dynamical response of the system
correctly. We derived them here just to facilitate the comparison with
the previous work on the subject. The correct expressions for ${\rm Im}\,
\Pi$ and $\sigma_{xx}$ are given in Sec.~\ref{L_scattering}.


\end{multicols}

\begin{references}
\bibitem{Comment_on_W_and_B} $W_{L, T}^2(r)$ should not be confused with
quantities $B_{L, T}(r)$ introduced in Ref.~\onlinecite{Giamarchi_95}.
For $W_L \ll W_T$, $B_L \simeq B_T \simeq W_T^2 / 2$.
\bibitem{Comment_on_Hall} Incidentally, this means that in the $\omega
\to 0$ limit, the collective-mode contribution to the magnetotransport
has the same form as in the ``Hall insulator,'' i.e., vanishing
$\sigma_{x x}$ but classical $\rho_{x y}$. However, at finite $T$ and
sufficiently low $\omega$ the single-particle contribution to the
transport will become more important. Therefore, it would be
unwarranted to draw the conclusion that the pinned WC behaves as a
Hall insulator just on the basis of the phonon response studied here.
\end{references}
\end{document}